\documentclass[sn-mathphys-ay]{sn-jnl}

\usepackage{graphicx}%
\usepackage{multirow}%
\usepackage{amsmath,amssymb,amsfonts}%
\usepackage{amsthm}%
\usepackage{mathrsfs}%
\usepackage[title]{appendix}%
\usepackage{xcolor}%
\usepackage{textcomp}%
\usepackage{manyfoot}%
\usepackage{booktabs}%
\usepackage{algorithm}%
\usepackage{algorithmicx}%
\usepackage{algpseudocode}%
\usepackage{listings}%

\usepackage{color,soul}
\usepackage{hyperref}
\usepackage{url}
\usepackage[left]{lineno}

\newcommand {\beq} {\begin{equation}}
\newcommand {\eeq} {\end{equation}}
\newcommand{\pr}[1]{\left(#1\right)}
\newcommand{\br}[1]{\left[#1\right]}
\newcommand{\norm}[1]{\left\lVert#1\right\rVert}
\newcommand{\abs}[1]{\left|#1\right|}

\begin{document}

\title[A model for contractile stress fibers embedded in bulk actomyosin networks]{A model for contractile stress fibers embedded in bulk actomyosin networks}

\author*[1]{\fnm{Mariya} \sur{Savinov} \orcid{https://orcid.org/0000-0002-5801-5710} }\email{mas10009@nyu.edu}

\author[1]{\fnm{Charles S.} \sur{Peskin }\orcid{https://orcid.org/0000-0003-3749-9864} }\email{peskin@cims.nyu.edu}

\author*[1,2]{\fnm{Alex} \sur{Mogilner} \orcid{https://orcid.org/0000-0002-9310-3812} }\email{mogilner@cims.nyu.edu}

\affil*[1]{\orgdiv{Courant Institute of Mathematical Sciences}, \orgname{New York University}}

\affil[2]{\orgdiv{Department of Biology}, \orgname{New York University}}

\abstract{Contractile cytoskeletal structures such as fine actomyosin meshworks and stress fibers are essential force-generators for mechanical phenomena in live cells, including motility, morphogenesis, and mechanosensing. While there have been many studies on the rheology and assembly of individual stress fibers, few mathematical models have explicitly modeled the bulk actomyosin network in which stress fibers are embedded, particularly not in the case of high actin turnover. Generally the extent of the interplay between embedded stress fibers and contractile bulk networks is still not well understood. To address this gap, we design a model of stress fibers embedded in bulk actomyosin networks which utilizes the immersed boundary method, allowing one to consider various stress fiber rheologies in the context of an approximately viscous, compressible, contractile bulk network. We characterize the dynamics of bulk actomyosin networks with and without embedded stress fibers, and simulate a laser ablation experiment to demonstrate the effective long-range interactions between stress fibers as well as how perturbations of stress fibers can result in symmetry breaking of the bulk actomyosin network.}


\maketitle

\section{Introduction}\label{Introduction}
Essential cellular processes such as motility, morphogenesis, and mechanosensing, among others, depend on the action of force-generating subcellular structures \citep{ruppel_force_2023}. Composed of actin, myosin motor proteins, and other complimentary components, these structures include the cytokinetic ring in cell division \citep{vavylonis_assembly_2008}, lamellipodia in motile cells \citep{cramer_organization_1999}, and stress fibers (SF) \citep{lehtimaki_generation_2021}. SFs are composed of bundles of cross-linked actin filaments \citep{cramer_identification_1997}, spanning up to tens of microns in length \citep{livne_inner_2016} with variable widths from less than $0.5\mu m$ \citep{livne_inner_2016} to potentially up to $1\mu m$ \citep{buenaventura_intracellular_2024}, typically thick and stable in non-motile cells while thinner and more dynamic in highly motile ones \citep{tojkander_actin_2012}. There are various classifications of SFs, particularly in the context of motile cells, outlined in reviews such as \cite{pellegrin_actin_2007, tojkander_actin_2012, kassianidou_geometry_2017}. Many types of SFs are contractile \citep{isenberg_cytoplasmic_1976,kreis_stress_1980}, containing myosin motor proteins, which convert ATP chemical energy to mechanical energy, generating forces and movement of actin filaments ultimately leading to a contractile tension along the length of the fiber. SFs additionally often have focal adhesions at their ends, connecting them to the extracellular environment. The contractile tension and adhesions of SFs are harnessed by the cell for a number of functions, such as regulation of nuclear shape \citep{khatau_perinuclear_2009}, persistent lamellipodial migration \citep{rid_last_2005}, resistance to shear in endothelial cells \citep{sato_biorheological_2005}, and mechanosensing \citep{hayakawa_actin_2008, colombelli_mechanosensing_2009,trichet_evidence_2012}.\\

There have been many studies on the rheology and assembly of individual SFs, both experimental \citep{cramer_identification_1997, bershadsky_assembly_2006, hotulainen_stress_2006,kumar_viscoelastic_2006, lu_mechanical_2008, russell_sarcomere_2009, tanner_dissecting_2010, tojkander_molecular_2011, lee_actomyosin_2018} and computational \citep{stachowiak_recoil_2009, russell_sarcomere_2009, besser_viscoelastic_2011, chapin_mathematical_2014, fogelson_actin-myosin_2018}. The response of individual SFs to laser ablation is well characterized to be viscoelastic \citep{kumar_viscoelastic_2006,tanner_dissecting_2010,besser_viscoelastic_2011, lee_actomyosin_2018}, and many have studied the dynamics by modeling SFs as active, Kelvin-Voigt-like structures \citep{besser_viscoelastic_2011, chapin_mathematical_2014, kassianidou_geometry_2017, bernal_actin_2022}. Few mathematical models of SFs explicitly model the surrounding cytoskeleton and cytoplasm in which SFs are immersed, assuming that they contribute an external drag force or to the internal fiber viscosity in a way which is homogeneous along the length of the fiber and independent of positioning in the cell. Dynamics relating to SF bending resistance are generally neglected, with models focusing instead on length-contraction dynamics.\\

However, experiments have shown that SFs are connected to and influenced by the bulk actomyosin network in cells \citep{vignaud_stress_2021,riedel_positioning_2024}, as well as other SFs \citep{kassianidou_geometry_2017}. For example, \cite{vignaud_stress_2021} captured the dynamics of ablated SFs in micropatterned cells by modeling them as contractile, elastic structures embedded in an elastic and percolated cortical network. Recently, \cite{riedel_positioning_2024} computationally modeled SFs embedded in an elastic bulk medium to explain SF positioning and its relation to cellular mechanical stress. These bulk actomyosin networks are also force-generating higher-order actin structures, where myosins produce contractile forces which cells harness for essential processes such as motility \citep{blanchoin_actin_2014} and morphogenesis \citep{heisenberg_forces_2013}. Though at the micron scale these networks are composed of semiflexible actin filaments which interact with each other through both proteins and hydrodynamic interactions, at the scale of tens of microns it is well understood that percolated actin networks behave viscoelastically, elastic on short timescales and viscous on long timescales \citep{schmoller_structural_2009, maxian_simulations_2021, maxian_interplay_2022}. The high turnover rates of \emph{in vivo} actomyosin networks \citep{brieher_mechanisms_2013} can make these networks effectively viscous on the timescale of minutes \citep{malik-garbi_scaling_2019}.\\

The extent to which embedding SFs in active, bulk actomyosin networks impacts cell mechanics is still not well understood. Moreover, how SF dynamics in, e.g., laser ablation experiments are affected by embedding in bulk actomyosin networks has not yet been studied through mathematical modeling. To address this gap, we design a model of SFs embedded in bulk actomyosin networks which utilizes the immersed boundary method (IBM) \citep{peskin_flow_1972, peskin_immersed_2002}. The  IBM has been used to study fluid-structure interactions in a variety of biological contexts, such as swimming \citep{fauci_computational_1988, tytell_interactions_2010, hamlet_numerical_2011, park_flagellated_2019}, biofilms \citep{dillon_modeling_1996}, bleb formation \citep{strychalski_computational_2013}, and cell migration \citep{strychalski_poroelastic_2015, lee_role_2017}. As a computational method, the IBM is advantageous in contexts where one seeks to consider various immersed boundary rheologies. In our utilization of the IBM, the immersed boundaries are SFs and the ``fluid" is the bulk actomyosin network. Though it is possible to model these networks on the microscopic level \citep{vavylonis_assembly_2008,maxian_simulations_2021,rutkowski_discrete_2021,maxian_interplay_2022, yan_toward_2022}, discretely and stochastically, such approaches are computationally intractable at the scale of tens of microns. Continuum models have been widely used to study the dynamics of bulk actomyosin networks \citep{lewis_actin-myosin_2014, linsmeier_disordered_2016, malik-garbi_scaling_2019, le_goff_actomyosin_2020}. So, we model the network as a compressible viscous fluid, matching experimental observations in cases of the high turnover characteristic of cells \citep{brieher_mechanisms_2013,malik-garbi_scaling_2019}. Myosin motors bind and unbind from the network, advecting or diffusing depending on their respective state. We present our model for SFs as 1D viscoelastic contractile structures, and use an immersed boundary approach to couple SFs to the bulk actomyosin network. Finally, we present results for dynamics of the bulk actomyosin network with and without embedded SF dynamics in 2D, to characterize the behaviors and inform parameter choices and applications in 3D. Our model highlights the interplay and balance between SFs and a contractile bulk actomyosin network, and the effective hydrodynamic interactions between different SFs across cellular-scale domains. We also show that laser ablation of SFs embedded in contractile bulk networks with fast turnover can yield a symmetry breaking effect.

\section{Bulk actomyosin network model}\label{sec:BulkNetworkModel}
\begin{figure}
\begin{center}
\includegraphics[width=1.0\linewidth, angle=0]{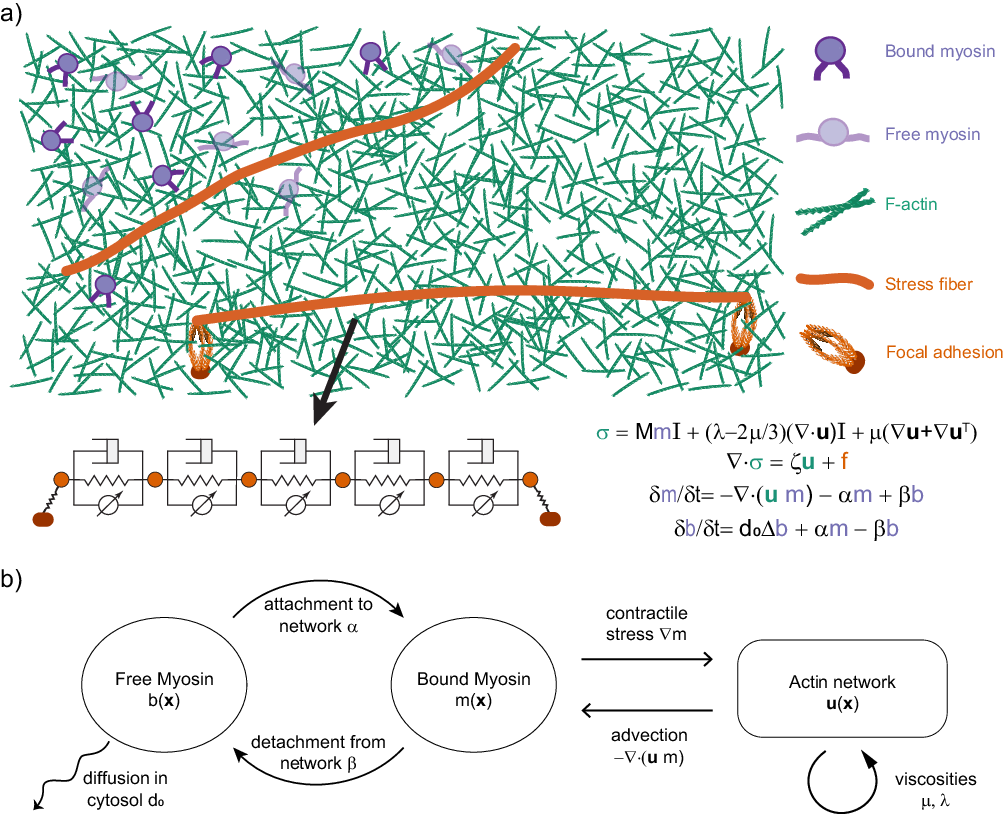}
\caption{\textbf{(a)} An illustration of the bulk actin network (green F-actin filaments), bound and free myosins (purple), embedded SFs (orange), and possible focal adhesions (orange). The contractile Kelvin-Voigt model for SFs is shown, as well as the bulk actomyosin force balance, reaction-advection, and reaction-diffusion equations. \textbf{(b)} A schematic of how the bound myosin, free myosin, and actin network interact with one another.}
\label{fig:Schematic}
\end{center}
\end{figure}
In Fig.~\ref{fig:Schematic}(a), we show an illustration of our system of a bulk actomyosin network with embedded SFs. We model the bulk actomyosin network as a viscous, compressible fluid with velocity $\pmb{u}\pr{\pmb{x}}$ which is driven by active compressive stresses due to a myosin density $m\pr{\pmb{x}}$. The stress tensor takes the form
\beq
\pmb{\sigma} = \br{M\,m+\pr{\lambda-2\mu/3}\nabla\cdot\pmb{u}}\pmb{I} + \mu\pr{\nabla\pmb{u}+\nabla\pmb{u}^T}
\label{eq:FluidStress}
\eeq
Here, stress due to myosin is isotropic, like a negative pressure on the fluid. The constants $\mu$ and $\lambda$ are the viscosity and expansion viscosity of the fluid, and $M$ is effectively the strength of myosin contraction. The expansion viscosity $\lambda$ must by positive by the second law of thermodynamics; for dilute monatomic gases, it is identically zero, but for water $\lambda\approx 3\mu$ \citep{cramer_numerical_2012}. Regardless, generally, the effect on the flow of the expansion viscosity term in \eqref{eq:FluidStress} tends to be small, but has been shown to be substantial in extreme cases such as when $\lambda \gg \mu$ in large Reynolds number flow \citep{cramer_effect_2014}.\\

At the cellular scale, the Reynolds number is small, close to $10^{-5}$ \citep{mogilner_intracellular_2018}, so there is effectively no inertia \citep{malik-garbi_scaling_2019}. Thus the network evolves by a balance of internal forces, defined by stress $\pmb{\sigma}$ in \eqref{eq:FluidStress}, with external forces. We assume, following previous models \citep{rubinstein_actin-myosin_2009}, that the network is subject to drag forces $\zeta \pmb{u}$ with the environment (in this case the cytosol and/or underlying surface) and forces $ \pmb{f}$ from any immersed boundaries (in our case, SFs). This yields a governing equation
\beq
\begin{split}
\nabla\cdot\pmb{\sigma} & = \pmb{f}_{\text{external}}\quad\quad\\
\underbrace{M\,\nabla m}_{\text{myosin}} + \underbrace{\pr{\lambda+\mu/3}\nabla\pr{\nabla\cdot\pmb{u}} + \mu \Delta\pmb{u}}_{\text{viscosity}} & = \underbrace{\zeta\,\pmb{u}}_{\text{drag}}+ \underbrace{\pmb{f}}_{\text{fiber}}\quad\quad
\end{split}
\label{eq:FluidForceBalanceDimensions}
\eeq
Throughout the paper, we assume that the actin network occupies a simply connected domain $\Omega$, in which \eqref{eq:FluidForceBalanceDimensions} applies. Often, the boundaries of such networks are free, but we will consider an equally ubiquitous case of a fixed domain where the high turnover of the network means the network quickly reassembles at the boundary to keep the domain fixed. We will assume, again following many previous models, that a no-stress boundary condition holds on this domain boundary:
\beq
\pmb{\sigma}\cdot\pmb{\hat{n}} = \pmb{0} \quad\text{on the boundary $\partial\Omega$}
\label{eq:FluidBC}
\eeq

Note that we do not consider a mass-conservation equation for the actin network itself, which would be an advection-reaction equation such as, for example:
\[
\dfrac{\partial \rho}{\partial t} = -\nabla\cdot\pr{\pmb{u}\rho} - \delta_d \,\rho + \delta_a
\]
The sample reaction terms here represent disassembly (with rate $\delta_d$) and assembly (with rate $\delta_a$) of the actin network. Generally, the viscosities ${\mu}$ and $\lambda$ may then depend on the density $\rho$ of the actin network, with denser networks being more viscous. However, as we are considering \textit{in vivo} dynamics, turnover of actin is generally fast and the reaction terms dominate over advection, eliminating variances in the apparent viscosity. For this reason, we do not consider the mass-conservation equation for the actin network.\\

In \eqref{eq:FluidForceBalanceDimensions}, the myosin density $m(\pmb{x})$ represents the fraction of myosin which is bound to the actin network and is advected by the network. However, myosin attachment to actin is dynamic, as myosin motors attach and detach stochastically at the molecular level. At the continuum level, this is reflected in there being two myosin populations in bulk: bound myosin $m$ and free myosin $b$, which we assume exchange with rates $\alpha$ and $\beta$ \citep{rubinstein_actin-myosin_2009}. While the bound myosin is advected by the network, we assume that the free myosin effectively freely diffuses in the cytosol. Thus, we model these two population dynamics by coupled reaction-advection and reaction-diffusion equations:
\beq
\begin{split}
\dfrac{\partial m}{\partial t} & = -\nabla\cdot \pr{\pmb{u}m} -\alpha\,m + \beta\,b\\
\dfrac{\partial b}{\partial t} & = d_0\,\Delta b + \alpha\, m - \beta\, b
\end{split}
\label{eq:Myosin}
\eeq
Here, $d_0$ is the diffusion coefficient for the free myosin, $\alpha$ is the detachment rate, and $\beta$ is the attachment rate. We assume effectively no-flux boundary conditions on the two myosin populations:
\beq
\begin{split}
m = 0, & \quad \pmb{x}\in\partial\Omega \quad \text{if} \quad \pmb{u}\cdot\pmb{n}<0\\
\dfrac{\partial b}{\partial n}=0, &\quad \pmb{x}\in\partial \Omega
\end{split}
\label{eq:MyosinBC}
\eeq
Here, $\pmb{n}$ is the outward facing normal of $\Omega$. The first part of \eqref{eq:MyosinBC} describes an inflow boundary condition on the bound myosin $m$ which enforces that no myosin flows into the domain $\Omega$. Note that there is no outflow condition because myosin generates contractile stresses and, in simulations with nonzero myosin densities, the network flow velocity is always directed inward.\\

The myosin dynamics equations \eqref{eq:Myosin} together with the force balance \eqref{eq:FluidForceBalanceDimensions} and boundary conditions \eqref{eq:FluidBC} and \eqref{eq:MyosinBC} effectively model the behavior of a bulk actomyosin network. Fig.~\ref{fig:Schematic}(b) shows a schematic of the action of the various parameters and components of the bulk actomyosin network. In the simulations, we will consider a fixed 2D rectangular domain $\Omega$. 

\subsection{Scaling and non-dimensionalization}\label{sec:BulkNetworkNondimensionalization}
Given a spatial scale $R$ (domain size) and myosin density scale $m_0$, we non-dimensionalize the actomyosin variables by
\[
\pmb{x}\rightarrow R\,\pmb{x}\quad\quad \pmb{u}\rightarrow \pr{\dfrac{Mm_0}{\zeta R}}\pmb{u}\quad\quad m\rightarrow m_0\, m\quad\quad b\rightarrow m_0\, b \quad\quad \pmb{f}\rightarrow Mm_0R\,\pmb{f}
\]
This then yields the fully nondimensional force-balance equation
\beq
\nabla m + \pr{\lambda+\mu/3}\nabla\pr{\nabla\cdot\pmb{u}} + \mu\Delta\pmb{u} = \pmb{u}+ \pmb{f}
\label{eq:FluidForceBalance}
\eeq
with two non-dimensional viscosity constants $\mu$ and ${\lambda}$ which relate to their dimensional counterparts by:
\[
\dfrac{\mu}{\zeta R^2}\rightarrow \mu \quad\quad \dfrac{\lambda}{\zeta R^2}\rightarrow \lambda
\]

There is a very useful notion of hydrodynamic length $\ell_0$, which relates to viscosity and drag by $\ell_0=\sqrt{\mu/\zeta}$, that shows how far in space a concentrated force spreads \citep{rubinstein_actin-myosin_2009}. The nondimensional viscosity coefficient $\mu$ in \eqref{eq:FluidForceBalance} is identical to the ratio of the hydrodynamic length to the domain scale squared: $\mu = (\ell_0/R)^2$.
If $\mu$ is large, then the hydrodynamic length is large relative to the domain size, and vice versa if $\mu$ is small.\\

Meanwhile, the myosin density equations \eqref{eq:Myosin} nondimensionalize as follows:
\[
\begin{split}
\dfrac{\partial m}{\partial t} & = -\nabla\cdot \pr{\pmb{u}m} -\alpha\,m + \beta\,b\\
\dfrac{\partial b}{\partial t} & = d_0\,\Delta b + \alpha\, m - \beta\, b
\end{split}
\]
The three, now non-dimensional, parameters relate to their dimensional counterparts by:
\[
 \dfrac{\zeta R^2}{M m_0}\alpha\rightarrow \alpha \quad\quad \dfrac{\zeta R^2}{M m_0}\beta\rightarrow \beta \quad\quad \dfrac{\zeta}{M m_0} d_0\rightarrow d_0
\]
where the left-side is in dimensions. Note that $\zeta R^2/M m_0$ is the characteristic timescale determined by the time of contraction driven by the balance of myosin-generated force and effective external drag.

\section{Stress fiber model}\label{sec:StressFiberModel}
Consider the SF as a 1D structure defined by Lagrangian points $\pmb{X}(q)$, where $q\in\Gamma$ denotes the position along the fiber $\Gamma$. There is a net internal force $\pmb{F}\pr{q}$ along the length of the fiber. In the case of low Reynolds numbers, the internal force is equal and opposite to the external force on the fiber from the network, which in turn is equal and opposite to the force on the network from the fiber:
\[
\pmb{F} = - \pmb{F}_{\text{external}}= - \pmb{F}_{\text{force on fiber from network}} = \pmb{F}_{\text{force on network from fiber}}
\]
This allows us to take advantage of an immersed boundary perspective, which typically relies on the immersed boundary being mass-less: the additional force $\pmb{f}$ we see in \eqref{eq:FluidForceBalanceDimensions} is related to the internal force $\pmb{F}\pr{q}$ in the fiber as:
\beq
\pmb{f}\pr{\pmb{x},t}=\int \pmb{F}\pr{q}\delta\pr{\pmb{x}-\pmb{X}\pr{q,t}}\,dq
\label{eq:FluidFiberForceContinuous}
\eeq
As such, given information as to the configuration of the fiber, one can solve just the bulk network equation \eqref{eq:FluidForceBalanceDimensions}, as opposed to two coupled equations for the network and fiber independently. We assume a no-slip boundary condition, as opposed to considering a formulation where the fibers move according to a slip velocity and contribution from the background bulk actomyosin network flow, such as \cite{kumar_cell_2015} do for blood flow or \cite{saintillan_theory_2015} for active particles. Our choice is motivated by previous experimental and modeling studies, which have shown that SFs are connected and embedded in the bulk actomyosin network in cells \citep{vignaud_stress_2021,riedel_positioning_2024}. Considering that SFs and the bulk networks are composed of the same building blocks, it is likely they may even interact directly through their complementary proteins, making a no slip boundary condition more relevant in this case. So, the SF (i.e. the immersed boundary) simply moves with the fluid velocity $\pmb{u}\pr{\pmb{x},t}$:
\beq
\dfrac{\partial \pmb{X}}{\partial t}\pr{q,t} = \int_{\Omega}\pmb{u}\pr{\pmb{x},t}\delta\pr{\pmb{x}-\pmb{X}\pr{q,t}}\,d\pmb{x}
\label{eq:FiberMotionContinuous}
\eeq

We will consider five classes of forces along the fiber: elastic forces $\pmb{F}^e$ due to fiber resistance to stretch/compression, bending forces $\pmb{F}^b$ due to fiber resistance to bending, viscous forces $\pmb{F}^v$ due to fluid-like fiber behaviors, myosin forces $\pmb{F}^m$ due to myosin activity, and finally adhesion forces $\pmb{F}^a$ due to adhesions at the SF ends. In all,
\[
\pmb{F}(q) = \pmb{F}^e(q) + \pmb{F}^b(q) + \pmb{F}^v(q) + \pmb{F}^m(q) + \pmb{F}^a(q)
\]
A schematic of the model can be seen in Fig.~\ref{fig:Schematic}(a), with the exception of the bending resistance response.

\subsection{Elastic forces}
Consider an elastic fiber with rest length $L_0$. It has some continuous elastic energy due to stretch or compression given by
\beq
E_e\br{\pmb{X}(q)} = \int_0^{L_0}\dfrac{K_E}{2}\pr{\norm{\dfrac{\partial \pmb{X}}{\partial q}}-1}^2\,dq
\label{eq:FiberEnergyElasticContinuous}
\eeq
The tension associated with this energy is
\beq
{T} = K_E\pr{\norm{\dfrac{\partial \pmb{X}}{\partial q}}-1}
\label{eq:FiberTensionContinuous}
\eeq
which is in the direction $\pmb{\tau}$, the unit tangent vector:
\beq
\pmb{\tau} = \dfrac{\partial \pmb{X}/\partial q}{\norm{\partial \pmb{X}/\partial q}}
\label{eq:FiberUnitTangentContinuous}
\eeq
The elastic force density on the fiber is then
\beq
\pmb{F}^e(q) = -\dfrac{\delta E_e}{\delta \pmb{X}} = \dfrac{\partial}{\partial q}\pr{T\pmb{\tau}}
\label{eq:FiberForceElasticContinuous}
\eeq
Note that the elastic energy of the fiber is in terms of some $K_E$ parameter with units of force, unlike a traditional spring constant, which has units of force over length. $K_E$ is essentially the Young's modulus of the fiber -- see Appendix~\ref{sec:KEYoungsModulus} for details. 

\subsection{Bending forces}
Suppose the fiber has some bending resistance, meaning that curvature has an associated energy given by
\beq
E_b\br{\pmb{X}(q)} = \int_0^{L_0}\dfrac{K_B}{2}\norm{\dfrac{\partial^2 \pmb{X}}{\partial q^2}}^2\,dq
\label{eq:FiberEnergyBendingContinuous}
\eeq
The elastic force density on the fiber is then
\beq
\pmb{F}^b(q) = -\dfrac{\delta E_b}{\delta \pmb{X}} = -K_B \dfrac{\partial^4 \pmb{X}}{\partial q^4}
\label{eq:FiberForceBendingContinuous}
\eeq
As this variational derivative of the bending energy is a fourth derivative of the configuration, or second derivative of the curvature, this is a particularly stiff term when solving this system numerically.

\subsection{Viscous forces}
It was shown that SFs exhibit viscoelastic behavior \citep{kumar_viscoelastic_2006, tanner_dissecting_2010,besser_viscoelastic_2011,lee_actomyosin_2018}. The nature of this viscoelasticity, however, depends on context and varies between systems. Previous research has shown that fibers in vitro behave like Kelvin-Voigt solids, which are elastic on long timescales but resistant to fast deformation \citep{kumar_viscoelastic_2006,stachowiak_recoil_2009}. Such materials are schematically represented as a spring and viscous element in parallel, with a Young's modulus $Y$ and viscosity $\eta$. Supposing a 1D fiber with rest length $L_0$ is stretched to length $L$, the strain $\varepsilon(t)=\pr{L-L_0}/L_0$ yields stress
\[
\begin{split}
\sigma(t) & = \eta \dfrac{d\varepsilon}{dt}+Y\,\varepsilon(t)
= \dfrac{\eta}{L_0} \dfrac{dL}{dt} + {Y}\pr{\dfrac{L}{L_0}-1}
\end{split}
\]
For fiber cross-sectional area $A$, the corresponding force is
\[
\begin{split}
F = A\sigma = & \dfrac{\eta\,A}{L_0}\dfrac{dL}{dt} + \dfrac{YA}{L_0}L
= \dfrac{\xi}{L_0} \dfrac{dL}{dt} + {K_E}\pr{\dfrac{L}{L_0}-1}
\end{split}
\]
Notice the second term is tension \eqref{eq:FiberTensionContinuous}, but for the case of a single spring. The first term above is the viscous force determined by deformation rate, with viscous coefficient $\xi=\eta A$. The viscous force on the fiber, in our coordinate system, is then given as
\beq
\pmb{F}^v = \xi \dfrac{\partial}{\partial t} \norm{\dfrac{\partial \pmb{X}}{\partial q}} \pmb{\tau}
\label{eq:FiberForceViscousContinuous}
\eeq
where the tangent vector \eqref{eq:FiberUnitTangentContinuous} defines the direction of the force.

\subsection{Myosin forces}
Many previous SF models \citep{kaunas_sarcomeric_2010,besser_viscoelastic_2011,kassianidou_geometry_2017} consider a linearized force-velocity relation for the myosin-generated contractile force. This relation is written as
\beq
F^m = F_s \pr{1+\dfrac{1}{v_0 L_0}\dfrac{dL}{dt}}
\label{eq:FiberForceMyosinContinuous}
\eeq
Here, $F_s$ is the stall force and $v_0$ is the free velocity {\it per unit length}. Namely, if the fiber is not deforming, it generates force (tension) $F_s$ along its length; if each fiber's unit contracts with free velocity, then myosin do not generate any force. This force $F^m$ is parallel to the tangent direction $\pmb{\tau}$ along the fiber given in \eqref{eq:FiberUnitTangentContinuous}.

\subsection{Adhesion forces}
Many SFs have focal adhesions at the ends, so as the fiber contracts it must do so under tension from these focal adhesions. Some studies have also suggested that there are adhesions to the environment throughout the length of the fiber \citep{sherrard_daam_2021}. We can implement the effect of focal adhesions through the use of \emph{target points}. Suppose we have an adhesion at a point $\pmb{Z}_j$. We model adhesions as springs with rest length zero, serving to keep the fiber position $\pmb{X}_j$ close to $\pmb{Z}_j$, the site of the adhesion. The corresponding force of this adhesion is then simply
\beq
\pmb{F}_j^{a} = -k_{adh}\pr{\pmb{X}_j-\pmb{Z}_j}
\label{eq:FiberForceAdhesion}
\eeq
where the parameter $k_{adh}$ is the force per unit length of the adhesive spring. However, if the adhesions are uniformly along the length of the fiber, we instead consider discrete forces
\beq
\pmb{F}_j^{a} = -k_{adh}\dfrac{\pmb{X}_j-\pmb{Z}_j}{\Delta q}
\label{eq:FiberForceAdhesionAlongLength}
\eeq

\subsection{Scaling and non-dimensionalization}
We nondimensionalize the fiber forces using the same scales introduced in Sec.~\ref{sec:BulkNetworkNondimensionalization}, summarized here in Table~\ref{tab:Scales}.
\begin{table}[h]
\caption{Scales for non-dimensionalization}\label{tab:Scales}
\begin{tabular*}{\textwidth}{@{\extracolsep\fill}ccccc}
\toprule%
 Spatial & Velocity & Myosin & Timescale & Fiber Force \\
\midrule
$R$  & $\dfrac{M m_0}{\zeta R}$ & $m_0$ & $\dfrac{\zeta R^2}{M m_0}$  & $f_0= M m_0 R$ \\
\botrule
\end{tabular*}
\end{table}
For the fiber, both the reference configuration $q$ and current configuration $\pmb{X}$ are scaled by $R$, so ultimately we seek to get the non-dimensional analogues of fiber parameters $K_E$, $K_B$, $\xi$, $F_s$, $v_0$, and $k_{adh}$.\\

For the elastic force \eqref{eq:FiberForceElasticContinuous}, we can nondimensionalize the elastic modulus $K_E$ by the $f_0=M m_0 R$ force scale to get the non-dimensional tension:
\beq
T = \dfrac{K_E}{f_0}\pr{\norm{\dfrac{\partial \pmb{X}}{\partial q}}-1} \longrightarrow K_E\pr{\norm{\dfrac{\partial \pmb{X}}{\partial q}}-1}
\label{eq:FiberTensionContinuousNonDim}
\eeq
Note that $\partial \pmb{X}/\partial q$ is equivalent to its nondimensional counterpart. Similarly, the tangent vector is without dimensions, so applying the above tension to \eqref{eq:FiberForceElasticContinuous} yields the non-dimensionalized elastic force density:
\beq
\pmb{F}^e(q) = -\dfrac{\delta E_e}{\delta \pmb{X}} = \dfrac{1}{R}\dfrac{\partial}{\partial q}\pr{T\pmb{\tau}}
\label{eq:FiberForceElasticContinuousNonDim}
\eeq
Note that for the numerical implementation, we will consider forces as opposed to force densities, so the $1/R$ factor will be dropped.\\

For the bending force \eqref{eq:FiberForceBendingContinuous}, we nondimensionalize the bending modulus $K_B$ using the force scale $f_0$ and spatial scale $R$ to get the nondimensional bending force density:
\beq
\pmb{F}^b(q) = -\dfrac{1}{R}\dfrac{K_B}{f_0 R^2} \dfrac{\partial^4 \pmb{X}}{\partial q^4} \longrightarrow -\dfrac{1}{R} K_B \dfrac{\partial^4 \pmb{X}}{\partial q^4}
\label{eq:FiberForceBendingContinuousNonDim}
\eeq
As we will be utilizing forces as opposed to force densities in our numerical implementation, we rescale $K_B/f_0R^2\rightarrow K_B$ and the additional $1/R$ factor above will be dropped.\\

The viscous force \eqref{eq:FiberForceViscousContinuous} is straightforward to nondimensionalize by utilizing the scale for time $\zeta R^2/M m_0$ and force $f_0$ as follows:
\beq
\pmb{F}^v = \dfrac{\xi}{f_0} \dfrac{M m_0}{\zeta R^2} \dfrac{\partial}{\partial t} \norm{\dfrac{\partial \pmb{X}}{\partial q}} \pmb{\tau} = \dfrac{\xi}{\zeta R^3}  \dfrac{\partial}{\partial t} \norm{\dfrac{\partial \pmb{X}}{\partial q}} \pmb{\tau} \longrightarrow \xi  \dfrac{\partial}{\partial t} \norm{\dfrac{\partial \pmb{X}}{\partial q}} \pmb{\tau} 
\label{eq:FiberForceViscousContinuousNonDim}
\eeq
Note that here we have a nondimensional force, not force density, so we rescale $ \xi/\zeta R^3\rightarrow \xi$.\\

For the myosin force \eqref{eq:FiberForceMyosinContinuous}, it is straightforward to nondimensionalize the fiber force $F_s\rightarrow F_S/f_0$ and the free velocity per unit length $ v_0\, Mm_0/\zeta R^2\rightarrow v_0$:
\beq
F^m = \dfrac{F_s}{f_0} \pr{1+\dfrac{1}{v_0 {M m_0} L_0/{\zeta R^2}}\dfrac{dL}{dt}}\rightarrow F_s \pr{1+\dfrac{1}{v_0L_0}\dfrac{dL}{dt}}
\label{eq:FiberForceMyosinContinuousNonDim}
\eeq
Finally, for the adhesion forces, we nondimensionalize the spring constant as $k_{adh} R/f_0\rightarrow k_{adh}$. If the adhesions are uniformly along the length of the fiber, instead nondimensionalize $k_{adh}/f_0\rightarrow k_{adh}$.\\

Table~\ref{tab:Parameters} lists the system parameters, both for the actomyosin network and SF, as well as their units and non-dimensional counterparts.

\begin{sidewaystable}
\caption{Scales for non-dimensionalization}\label{tab:Parameters}
\begin{tabular*}{\textwidth}{@{\extracolsep\fill}cccc}
\toprule%
 Parameter name & Notation & Units & Non-dimensional analogue \\
\midrule \tabularnewline
Actin network viscosity (2D)  &  $\mu$ & force per length $\times$ time & $\mu_{n.d.}=\dfrac{\mu}{\zeta R^2}$ \tabularnewline\\ \hline \tabularnewline
    Actin network 2nd viscosity (2D)  &  $\lambda$ & force per length $\times$ time & $\lambda_{n.d.}=\dfrac{\lambda}{\zeta R^2}$ \tabularnewline\\ \hline \tabularnewline
    Actin network drag coefficient & $\zeta$ & force per length cubed $\times$ time & $\zeta_{n.d.}=1$\tabularnewline\\ \hline \tabularnewline
    Myosin stress coefficient & $M$ & force $\times$ length per myosin concentration & $M_{n.d.}=1$\tabularnewline\\ \hline \tabularnewline
    Elastic fiber force coefficient & $K_E$ & force & $K_{E,\,n.d.} = \dfrac{K_E}{f_0}$\tabularnewline\\ \hline \tabularnewline
    Bending fiber force coefficient & $K_B$ & force $\times$ length squared & $K_{B,\,n.d.}=\dfrac{K_B}{f_0 R^2}$\tabularnewline\\ \hline \tabularnewline
    Fiber viscosity & $\xi$ & force $\times$ time & $\xi_{n.d.}=\dfrac{\xi}{\zeta R^3}=\dfrac{\xi}{f_0}\dfrac{M m_0}{\zeta R^2}$\tabularnewline \\ \hline \tabularnewline
    Myosin stall force & $F_s$ & force & $F_{s, n.d.}=\dfrac{F}{f_0}$ \tabularnewline\\ \hline \tabularnewline
    Myosin free velocity & $v_0$ & velocity per unit length & $v_{0, n.d.}=\dfrac{\zeta R^2}{M m_0} v_0$\tabularnewline\\ \hline \tabularnewline
    Adhesion spring coefficient & $k_{adh}$ & force per unit length & $k_{adh, n.d.}=\dfrac{R}{f_0} k_{adh}$\tabularnewline\\ 
\botrule
\end{tabular*}
\end{sidewaystable}

\section{Immersed boundary method for a fiber embedded in an actomyosin network}
Consider now that the fiber is immersed in a global actin mesh, which behaves like the active compressible viscous fluid outlined in Sec.~\ref{sec:BulkNetworkModel}. Following the introduction in Sec.~\ref{sec:StressFiberModel}, let $\pmb{F}(q)$ define the force along the fiber (and thus the force the fiber exerts on the fluid). The traditional immersed boundary method (IBM) with no-slip boundary conditions has governing equations \eqref{eq:FluidForceBalance}, \eqref{eq:FluidFiberForceContinuous}, and \eqref{eq:FiberMotionContinuous}, restated here:
\begin{align*}
\text{Fluid force balance:}\quad& {k\,\nabla m} + {\pr{\lambda+\mu/3}\nabla\pr{\nabla\cdot\pmb{u}} + \Delta\pmb{u}}  = {\zeta\,\pmb{u}}+ {\pmb{f}} && \eqref{eq:FluidForceBalance}\\\\
\text{Force spreading:}\quad &  \pmb{f}\pr{\pmb{x},t}=\int \pmb{F}(q)\delta\pr{\pmb{x}-\pmb{X}\pr{q,t}}\,dq = \pmb{S}\pr{\pmb{X}}\pmb{F}(q) && \eqref{eq:FluidFiberForceContinuous}\\\\
\text{Fiber equation of motion:} \quad & \dfrac{\partial \pmb{X}}{\partial t}\pr{q,t} = \int_{\Omega}\pmb{u}\pr{\pmb{x},t}\delta\pr{\pmb{x}-\pmb{X}\pr{q,t}}\,d\pmb{x}=\pmb{S}^*\pr{\pmb{X}}\pmb{u} && \eqref{eq:FiberMotionContinuous}
\end{align*}
Here we define the spreading operator $\pmb{S}$, which has adjoint $\pmb{S}^*$ defining the interpolation operator. The two operators must be adjoint to one another if energy conservation is required \citep{peskin_immersed_2002}.\\

Previous studies have shown that immersed boundaries have some hydrodynamic radius $r_h$, associated with the discretized delta function which depends directly on the Eulerian grid size $h$ \citep{peskin_immersed_2002,bringley_validation_2008}. This hydrodynamical radius is typically of $O(h)$, such as $r_h \sim 1.2h-1.4h$ in the Stokes case \citep{bringley_validation_2008, maxian_immersed_2020}. So, in order to resolve the diameter of the body, the Eulerian grid size $h$ is set such that the physical radius $a$ of the fiber matches the hydrodynamic radius $r_h$. For cytoskeletal components, the physical radius $a$ can be very small, making $h$ prohibitively small. In traditional IBM, the Eulerian and Lagrangian grid sizes need to be comparable \citep{peskin_immersed_2002,bringley_validation_2008}, so if the Eulerian grid is small, the Lagrangian discretization size must also be small for stability. Thus, as a consequence of a small immersed boundary radius $a$, one must solve a stiff system of equations which requires a small timestep $\Delta t$ to resolve the behavior.\\

Stress fibers have variable widths, often smaller than $0.5\mu m$ \citep{livne_inner_2016} but could potentially be as large as a micron in width \citep{buenaventura_intracellular_2024}. In this work, we will apply our IBM formulation for the case of thicker SFs ($\approx 0.5 - 1\mu m$ in diameter) in thin actomyosin bulk networks spread on cellular scales ($\approx 50\mu m$ domain sizes). The 2D problem is applicable for example to questions regarding actomyosin and SF dynamics in thin cells spread on flat surfaces \citep{ruppel_force_2023}. In such cases, it is sufficient to consider grid sizes $h\sim 1/200- 1/100$ to simulate fibers of realistic thickness. In future work for 3D bulk networks, we can follow the approach of \cite{maxian_immersed_2020} to leverage an analytical result regarding the drag on a sphere immersed in the fluid, instead considering a fiber equation of motion given by
\beq
\dfrac{d\pmb{X}}{dt}(q,t) = \pmb{S}^*\pr{\pmb{X}}\pmb{u} + \dfrac{1}{\gamma}\pmb{F}(q)
\label{eq:FiberMotionContinuousSlip}
\eeq
where $\gamma$ is a function of the hydrodynamic radius $r_h$, the true radius of the fiber $a$, and the viscosities of the actin network $\mu$ and $\lambda$:
\beq
\gamma = 108\pi \dfrac{\mu\lambda\pr{3\lambda + 4\mu}}{\pr{3\lambda+10\mu}\pr{6\lambda+11\mu}} \dfrac{r_h a}{r_h-a} = 108\pi\eta \dfrac{r_h a}{r_h-a}
\label{eq:gamma}
\eeq
This modified equation of motion \eqref{eq:FiberMotionContinuousSlip} would then allow one to simulate the dynamics of thin SFs with computationally accessible coarser grid sizes, retaining the appropriate drag on the SFs. Moreover, this modified IBM formulation \eqref{eq:FiberMotionContinuousSlip} is advantageous in contexts where the focus is on bulk flows on larger scales than the fiber radius $a$. For details, see Appendix~\ref{sec:ModifiedIBM}.

\subsection{Temporal discretization}
Our spatial discretization of the spreading of the fiber force to the fluid \eqref{eq:FluidFiberForceContinuous} and the interpolation of the fluid velocity to the fiber \eqref{eq:FiberMotionContinuous} follows previous work, where the $\delta$-functions are approximated with the four-point delta function $\delta_{\Delta x,\Delta y}$, detailed in Appendix~\ref{sec:4ptDelta}. The force spreading \eqref{eq:FluidFiberForceContinuous} is discretized as 
\beq
\pmb{f}\pr{\pmb{x},t} = \sum_{k=1}^{N_b} \pmb{F}_k\pr{\pmb{X}_1,...,\pmb{X}_{N_b}}\delta_{\Delta x,\Delta y}\pr{\pmb{x}-\pmb{X}_k(t)}
\label{eq:FluidFiberForceDiscrete}
\eeq
and the interpolation of the fluid velocity onto the Lagrangian grid is discretized as
\beq
\br{\pmb{S}^*\pmb{u}}_k = \sum_{\pmb{x}} \pmb{u}\pr{\pmb{x},t} \delta_{\Delta x,\Delta y}\pr{\pmb{x}-\pmb{X}_k(t)}\Delta x\Delta y
\label{eq:FiberMotionDiscreteSlip}
\eeq
As the Eulerian grid is finitely large and the delta function is compactly supported, the sum above is finite in both components. The spatial discretization $\pmb{F}_k$ of the fiber forces $\pmb{F}(q)$ is outlined in Appendix~\ref{sec:FiberDiscretization}.\\

We solve the network force balance equation using a second-order finite difference scheme, as outlined in Appendix~\ref{sec:FluidScheme}, and denote the resultant discretized force balance equation as
\beq
\pr{\lambda+\mu/3}\pmb{D}\pr{\pmb{D}\cdot\pmb{u}} + \mu L\pmb{u} = -k\pmb{D}m + \zeta\pmb{u} + f_0\pmb{f}
\label{eq:FluidForceBalanceDiscrete}
\eeq
where $\pmb{D}$ is the discretized Del operator and $L$ is the discretized Laplacian.\\

In the case of a SF without velocity-dependent forces, it is straightforward to timestep, as the fiber configuration determines fiber force and thus bulk velocity, which interpolates to the fiber velocity for timestepping (see details in Appendix~\ref{sec:TraditionalTimestep}). However, for a contractile, viscoelastic fiber, the fiber forces depend on \emph{both} the current configuration $\pmb{X}$ and fiber velocity $\pmb{U}$: 
\[
\pmb{F} = \underbrace{\pmb{F}^e\pr{\pmb{X}} + \pmb{F}^a\pr{\pmb{X}}}_{\text{nonlinear in }\pmb{X}}+\underbrace{\pmb{F}^b
\pr{\pmb{X}}}_{\text{linear in }\pmb{X}}+ \underbrace{\pmb{F}^v\pr{\pmb{X},\pmb{U}}+\pmb{F}^m\pr{\pmb{X},\pmb{U}}}_{\text{nonlinear in $\pmb{X}$, linear in $\pmb{U}$}}
\]
As such, as opposed to the traditional IBM fiber equation of motion \eqref{eq:FiberMotionContinuous}, we have an implicit equation for the fiber velocity $\pmb{U}$:
\beq
\dfrac{\partial \pmb{X}}{\partial t} = \pmb{U} = \pmb{S}^*\pr{\pmb{X}}\pmb{u}\pr{\pmb{X},\pmb{U}}
\label{eq:FiberMotionContinuousModified}
\eeq
To determine the fiber configuration over time, we need to solve \eqref{eq:FiberMotionContinuousModified} with an iterative approach. We choose to timestep with a fixed point iteration algorithm, where we use the previous fiber velocity $\pmb{U}^{n-1}$ as an initial ``guess" for the current velocity $\pmb{U}^{n,(0)}$. Then, the updated fiber velocity $\pmb{U}^{n,(m)}$ and configuration $\pmb{X}^{n+1,(m)}$ at step $m$ of the iteration is given by 
\beq
\dfrac{\pmb{X}^{n+1,(m)}-\pmb{X}^n}{\Delta t} = \pmb{U}^{n,(m)} = \pmb{S}^*\pr{\pmb{X}^n}\pmb{u}\pr{\pmb{X}^n,\pmb{U}^{n,(m-1)}}
\label{eq:FiberMotionDiscreteModified}
\eeq
Once the relative change in either the fiber velocity or configuration reaches some tolerance, one takes the last output of the fixed point iteration as defining the updated velocity and configuration.\\

In principle, one could choose to only iterate at initial time $t=0$, and at later times take only $m=1$ iterative steps, which would correspond to time-lagging the velocity. Such an approach was used, for example, by \cite{strychalski_viscoelastic_2012} to simulate Kelvin-Voigt immersed structures in Stokes flow. However, in the instance of actions such as laser ablation of the SFs, where the forces would change dramatically due to tension release, it would be necessary to iterate to use a more accurate fiber velocity. In practice, we find the most expensive iterative step to be the first one, with later timesteps requiring far fewer iterations. As such, we choose to implement a tolerance condition at each timestep for consistency.\\

We find that our implementation yields a restriction on the internal viscosity of the fiber $\xi$ as it relates to the fluid viscosity $\mu$. We observe that, for a fixed Eulerian grid, there is typically some critical value $\xi/\mu$ above which the initial fixed point iteration does not converge. This is similar to the restriction found by \cite{strychalski_viscoelastic_2012}. In our simulations, we take the maximal value of $\xi$ for which our method converges.

\section{Results}
\subsection{Bulk actomyosin network dynamics}
The bulk actomyosin network dynamics are governed by five nondimensional parameters: the myosin detachment rate $\alpha$, the myosin attachment rate $\beta$, the free myosin diffusion coefficient $d_0$, the first fluid viscosity $\mu$ (which in its non-dimensional form is equivalent to the hydrodynamic length over the domain length squared), and finally the expansion fluid viscosity $\lambda$. Before considering an embedded SF, we seek to understand the behaviors of the bulk network in different parameter regimes.\\

One important feature of our model is the positive feedback between the bound myosin, which is advected by the actin network, and the actin network, which is driven by gradients in the bound myosin distribution. Consider the following scenario: there is a concentration of bound myosin in a region of characteristic size $\ell<R$ somewhere in the network. As a result, the network will start to contract, flowing inward around the region and thus causing its characteristic size to shrink. Smaller $\mu$ and $\lambda$ should correspond to faster contractile flows, as the viscous stress is smaller. If myosin remains bound to the network, the same mass of myosin now occupies a smaller region, so there is then a larger gradient of myosin and, as a result, a larger contractile force. If not for the exchange of bound and free myosin as well as network viscous resistance, this contractile positive feedback would continue, compressing the myosin into a very small region of the domain. Fig.~\ref{fig:Schematic}(b) depicts a schematic of this inherent interplay of factors.\\

\begin{figure}
\begin{center}
\includegraphics[width=1.0\linewidth, angle=0]{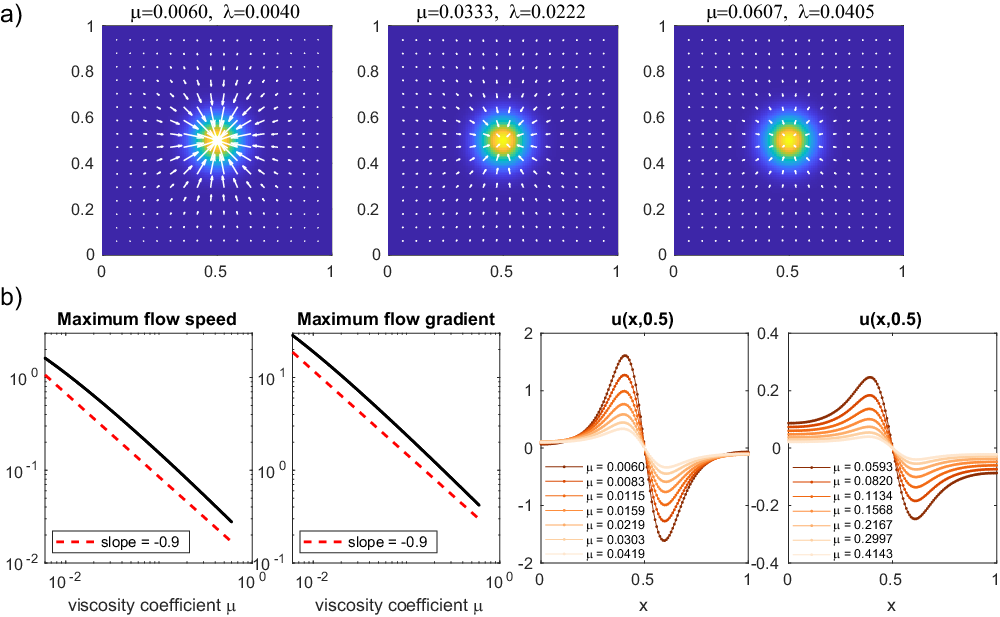}
\caption{\textbf{(a)} Numerical solutions for the flow of the actomyosin network for different network viscosities in the case of a gaussian-like distribution of bound myosin \eqref{eq:MyosinPeak} in the center of the domain $\Omega=[0,1]\times[0,1]$. The first viscosity coefficient $\mu$ is varied and the expansion viscosity $\lambda=2\mu/3$ in all cases. The flow quiver arrows are identically scaled between all subplots, such that a flow speed of $1$ appears as an arrow of size $0.07$ in the domain. \textbf{(b)} Quantitative differences in flow properties around the myosin distribution in (a) as a function of changing actin network first viscosity coefficient $\mu$ with expansion viscosity $\lambda=2\mu/3$. The left two columns show the maximum flow speeds and gradients, respectively, as a function of increasing viscosity on a log-log plot. The right two columns shows the flow distribution along the horizontal midline of the myosin peak as a function of horizontal position $x$ for a few cases of $\mu$, ranging from $\mu=0.006$ to $\mu=0.4143$, with lighter colors corresponding to larger values of $\mu$. All solutions are computed on a $128\times128$ Eulerian grid.}
\label{fig:FluidParamSweep}
\end{center}
\end{figure}

To investigate the role of parameters in the actomyosin network, consider a spot of bound myosin represented by a distribution:
\beq
m(x,y) = \exp\pr{-\dfrac{\pr{x-x_0}^2}{0.1^2}-\dfrac{\pr{y-y_0}^2}{0.1^2}}
\label{eq:MyosinPeak}
\eeq
Fig.~\ref{fig:FluidParamSweep}(a) shows the case of a spot located at the center $(x_0,y_0)=(0.5,0.5)$ of the non-dimensional domain (see Fig.~S1 for the case of a spot in the corner).\\

First, let us consider the impact of the actin network viscosity parameters $\mu$ and $\lambda$. Fig.~\ref{fig:FluidParamSweep}(a) shows the numerical solutions for actin network flows around a centered myosin spot for different orders of magnitude of $\mu$ and $\lambda = 2\mu/3$ (see Fig.~S1 for the corner case). Note that we take $\lambda=2\mu/3$ so that both viscosity terms in \eqref{eq:FluidForceBalance} have the same coefficient. All the flow vectors have the same scale, and it is clear that as the viscosity parameters increase (left to right) the flow speed decreases. We can quantify this by measuring the maximium flow speed and maximum flow gradient as $\mu$ increases, keeping $\lambda=2\mu/3$. Fig.~\ref{fig:FluidParamSweep}(b) shows the maximum flow speed and maximum flow gradients for varying $\mu$, and we see that the maximum flow speed and maximum flow gradient decrease approximately like $1/\mu^{0.9}$. The flow profile at $y=0.5$ as one varies the $x$ coordinate shows a centripetal-like flow, which decays as $\mu$ and $\lambda$ are increased (dark to light), thus featuring both a lower maximal contraction and consequentially a smaller flow gradient. Notice that Fig.~\ref{fig:FluidParamSweep} also shows that the flow speed and gradients are most sensitive for small $\mu$ and $\lambda$ (as the speed and gradient versus $\mu$ are on a log-log plot). Recall that the nondimensional viscosity is $\mu=\pr{\ell_0^2/R^2}$, where $R$ is the domain size and $\ell_0$ is the hydrodynamic length. One can interpret hydrodynamic length as follows: if $\ell_0$ is large, the actin network resists flow primarily due to internal friction rather than from drag with its environment. Effectively, the network is very stiff, and effects of small shear stresses are felt at long range, integrating both the internal friction and external drag over a large size. If on the other hand $\ell_0$ is small, resistance to flows is significantly more local. Thus, as $\mu$ is increasing, the effective hydrodynamic length $\ell_0$ increases as the network grows stiffer and forcing from myosin contraction is resisted at distances further and further away from the myosin spot. So intuitively, dynamics are more sensitive to changes in a small hydrodynamic length as compared to a large hydrodynamic length, as shown in the flow speed and flow gradient sensitivity to small $\mu$.\\

Note that we have fixed $\lambda=2\mu/3$ and varied $\mu$. Changing $\lambda$ does not result in qualitatively different flow profiles compared with increasing $\mu$ and $\lambda$ together, and similarly an increase in the expansion viscosity results in a slower flow speeds and smaller gradients. For details, see Fig.~S2.\\
\begin{figure}
\begin{center}
\includegraphics[width=1.0\linewidth, angle=0]{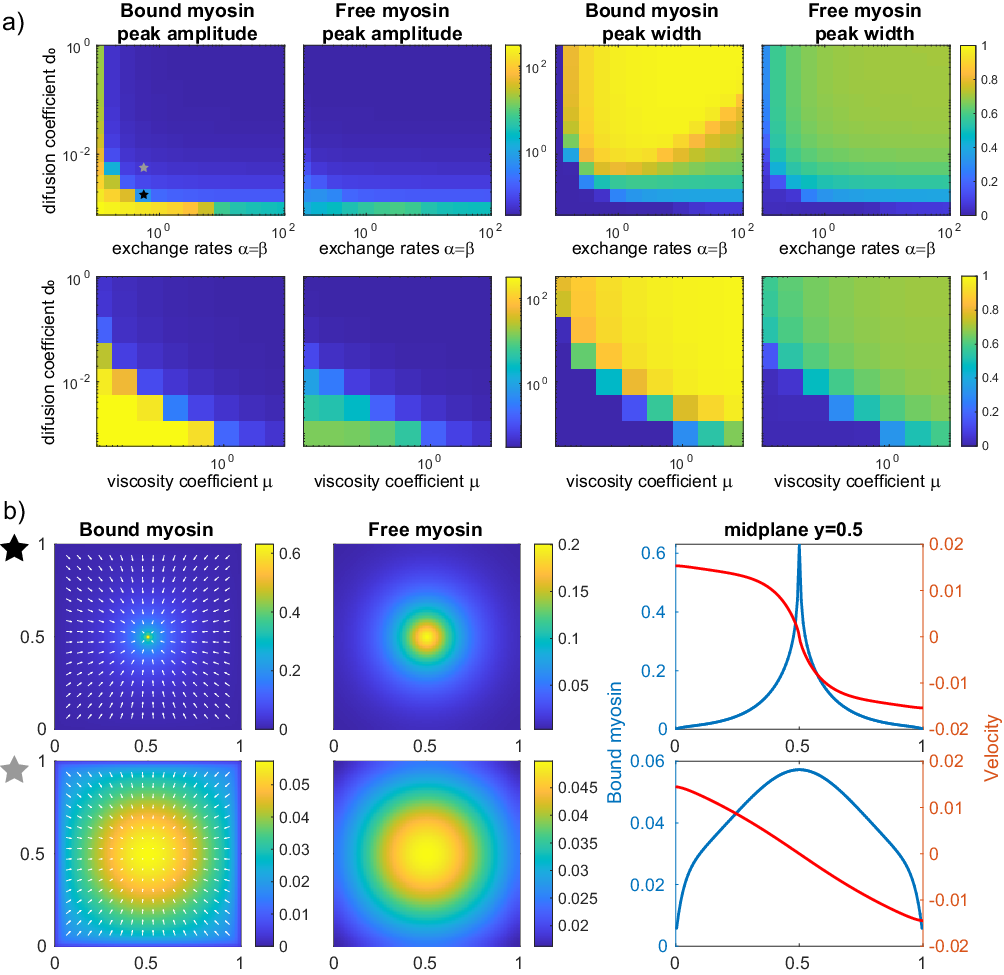}
\caption{\textbf{(a)} The peak amplitudes (left) and half-widths (right) for steady state distributions of bound and free myosin in the cases of varying the diffusion coefficient $d_0$ and reaction rates $\alpha=\beta$ for fixed $\mu=0.6$, $\lambda=0.4$ (top) and varying the diffusion coefficient $d_0$ and viscosity coefficient $\mu$, $\lambda=2\mu/3$ for fixed $\alpha=\beta=1$ (bottom). The myosin populations are initialized both with distribution \eqref{eq:MyosinPeak}, depicted in Fig.~\ref{fig:FluidParamSweep}(a). The steady state distributions are computed by numerically solving the system on a $128\times128$ grid with timestep $\Delta t = 0.001$ until the relative $L_{\infty}$ norm of the myosin distribution over $0.2$ time units varies by less than $10^{-4}$. There are minimal quantitative differences if a smaller grid-size and timestep are used. \textbf{(b)} The final steady state distributions of bound and free myosin when $\alpha=\beta=0.5623$, $\mu=0.6$, and $\lambda=0.4$ with $d_0=0.00178$ (top) and $d_0=0.00562$ (bottom).}
\label{fig:MyoFluidParamSweep}
\end{center}
\end{figure}

Of course, given such flows as in Fig.~\ref{fig:FluidParamSweep}, the bound myosin spots would in principle start to compact, and potentially migrate as in the case of a spot in the corner of the domain. What sort of myosin distributions are steady depends not only on the fluid viscosities $\mu$ and $\lambda$, but also on the myosin unbinding and binding parameters $\alpha$ and $\beta$, as well as the diffusion coefficient $d_0$ of the free myosin population. Recall the interplay of the various parameters is depicted schematically in Fig.~\ref{fig:Schematic} -- the diffusion of free myosin could act as a balance to the contraction and compaction of myosin, conditionally on the reaction rates $\alpha$ and $\beta$ being sufficiently large. However, it is unclear how large or small $\alpha$, $\beta$, and $d_0$ should be to achieve different steady distributions.\\

Consider the initial condition of a myosin spot in the center of the domain in which both the bound and free myosin have the same initial distribution \eqref{eq:MyosinPeak}, as depicted in Fig.~\ref{fig:FluidParamSweep}(a). Let us first fix the fluid viscosities $\mu=0.6$ and $\lambda=0.4$, as an example, and simulate the actomyosin system for different $d_0$ and $\alpha=\beta$ choices until the distribution of bound myosin stops changing within some tolerance, which we choose to be a $<10^{-4}$ difference in the relative $L_{\infty}$ norm over $0.2$ time units. Note that we keep $\alpha$ and $\beta$ equal, as we do not observe significant changes in the results when the binding and unbinding rates are of the same order of magnitude. Our aim is to assess the qualities of a `steady state myosin spot', in terms of the bound and free myosin maximal peak values and peak widths, from an initial starting point of maximum amplitude $1$ and width of approximately $0.1665$.\\

The first row of Fig.~\ref{fig:MyoFluidParamSweep}(a) shows the bound and free myosin final peaks and widths for different choices of reaction rates $\alpha=\beta$ and diffusion coefficient $d_0$. Generally, Fig.~\ref{fig:MyoFluidParamSweep}(a) shows that even relatively small diffusion coefficients are sufficiently large to smooth out myosin distributions so long as the exchange rates are fast enough (e.g. $\alpha=\beta>0.5$). Consider for example the case of $\alpha=\beta=0.5623$ for two choices of diffusion coefficient: $d_0=0.00178$ and $d_0=0.00562$. These two parameter sets are denoted by a black star and gray star in the first row of Fig.~\ref{fig:MyoFluidParamSweep}(a). Fig.~\ref{fig:MyoFluidParamSweep}(b) shows the steady state distributions for these two cases, where in the smaller diffusion coefficient case the bound myosin distribution is sharply peaked at the center with an amplitude of approximately $0.6$. Meanwhile, increasing the diffusion coefficient by roughly a factor of three results in a very smooth, nearly uniform myosin distribution, where the concentrated quality of the myosin is effectively lost. On the other hand, Fig.~\ref{fig:MyoFluidParamSweep}(a) also demonstrates that diffusion coefficient eventually becomes irrelevant if the exchange of bound and free myosin is small, in which case the bound myosin compacts very quickly to a peak of width $\ll 1$ and it does not unbind from the network fast enough to decrease the contraction speed. A similar result occurs if $d_0=0.001$, where even if the exchange rates are fast, the diffusion is too slow to redistribute the myosin fast enough to prevent compaction of the spot.\\

We also see in Fig.~\ref{fig:MyoFluidParamSweep}(a) that as $\alpha$ and $\beta$ become large, the bound and free myosin characteristics start to match, first for small diffusion $d_0$, and as $\alpha=\beta$ increases for larger diffusion coefficients. This is an effect of behavior in the limit of large exchange rates $\alpha$ and $\beta$, where the exchange of bound and free myosin is the fast process of \eqref{eq:Myosin} while the advection and diffusion terms are relatively slow. In this limit, $m \approx \beta b/\alpha$ and the total myosin distribution $M=m+b$ obeys a modified advection-diffusion equation:
\[
\dfrac{\partial M}{\partial t} = - \dfrac{\beta}{\alpha+\beta} \nabla\cdot\pr{\pmb{u}M} + \dfrac{\alpha}{\alpha+\beta} d_0 \Delta M
\]
So, Fig.~\ref{fig:MyoFluidParamSweep}(a) shows how as $\alpha=\beta$ becomes larger the bound and free myosin distributions approach $m\approx b\approx M/2$. This limit is reached faster for smaller diffusion coefficients $d_0$, where the exchange rates $\alpha$ and $\beta$ become the dominating terms more quickly.\\

The unbinding of myosin from the network and then diffusing outward is not the only process which balances the positive feedback of the advected myosin inducing contraction on the network. Recall Fig.~\ref{fig:FluidParamSweep}(b), which showed that as fluid viscosities $\mu$ and $\lambda$ increase, the maximum flow speed of the network decreases. In principle, a larger diffusion coefficient could still yield a sufficiently concentrated myosin distribution if the flow speeds were higher than those such as in Fig.~\ref{fig:MyoFluidParamSweep}(b). So, fixing $\alpha=\beta=1$, the bottom row of Fig.~\ref{fig:MyoFluidParamSweep}(a) shows how the bound and free myosin peak amplitudes and widths change when varying viscosity $\mu$ ($\lambda=2\mu/3$) and diffusion coefficient $d_0$. As expected, increasing diffusion or increasing the actin network viscosity yields similar effects on the network, where the bound myosin quickly transitions from a highly concentrated distribution to a relatively smooth and uniform distribution. The effect on the free myosin is analogous. Overall, Fig.~\ref{fig:MyoFluidParamSweep} shows that when it comes to myosin `spot'-like distributions, the actomyosin network tends to the extremes of highly concentrated or highly smoothed myosin distributions, with only fine-tuned parameters achieving a balance between the extremes.

\subsection{Isolated stress fiber dynamics}
Before embedding SFs into such actomyosin networks, we will first briefly investigate the dynamics of an isolated SF, where the interior forces are in balance with each other. Doing so, we aim to understand the impact of the five parameters determining SF rheology:  the elastic resistance $K_E$ (proportional to Young's modulus), bending resistance coefficient $K_B$,  viscous resistance $\xi$, contractile stall force $F_s$, and free velocity of myosin per unit length $v_0$.\\

If bending coefficient $K_B$ is small, the fast process is that of length change. Suppose there is no myosin activity, and let the change in length over time be given by  $\ell(t) = L_0-L(t)$, where $L_0$ is the rest length of the fiber. Approximately, in short times, this change in length obeys the ordinary differential equation
\[
\dfrac{\xi}{\L_0}\dfrac{d\ell}{dt} + \dfrac{K_E}{L_0}\ell = 0
\]
which has solution
\[
\ell(t) = \ell_0\exp\pr{-t/\tau_L}
\]
for an initial deformation $\ell_0$. Here, the timescale of length relaxation $\tau_L$ is given by
$\tau_L = \xi/K_E$.
Intuitively, a larger fiber viscosity coefficient $\xi$ leads to slower length relaxation, as the fiber is more resistant to fast deformation. Similarly a smaller Young's modulus $K_E$ also leads to slower length relaxation, as the fiber is less resistant to large deformations. If now we include contractile forces, there is an additional term in the differential equation for the change in length:
\[
\dfrac{\xi}{L_0}\dfrac{d\ell}{dt} + \dfrac{K_E}{L_0}\ell = F_s\pr{1-\dfrac{1}{v_0L_0}\dfrac{d\ell}{dt}}
\]
This contractile force contributes a forcing term as well as complements the viscous resistance as $\xi\rightarrow \xi+F_s/v_0$. The solution to the differential equation for the change in length $\ell$ over time, given some initial deformation $\ell_0$, is
\[
\ell(t) = \dfrac{F_s L_0}{K_E}\pr{1-\exp\pr{-t/\tau}} + \ell_0\exp\pr{-t/\tau}
\]
where the timescale of length relaxation is now 
\beq
\tau_L = \dfrac{\xi+F_s/v_0}{K_E}
\label{eq:tauL}
\eeq
Suppose the stall force $F_s$ is fixed -- as such, the final contractile length is fixed. This shows that smaller free velocities lead to slower length change, which makes sense physically as the free velocity relates to the rate at which myosin traverses actin. If myosin is slower, contraction is inherently slower.\\

\begin{figure}
\begin{center}
\includegraphics[width=0.95\textwidth, angle=0]{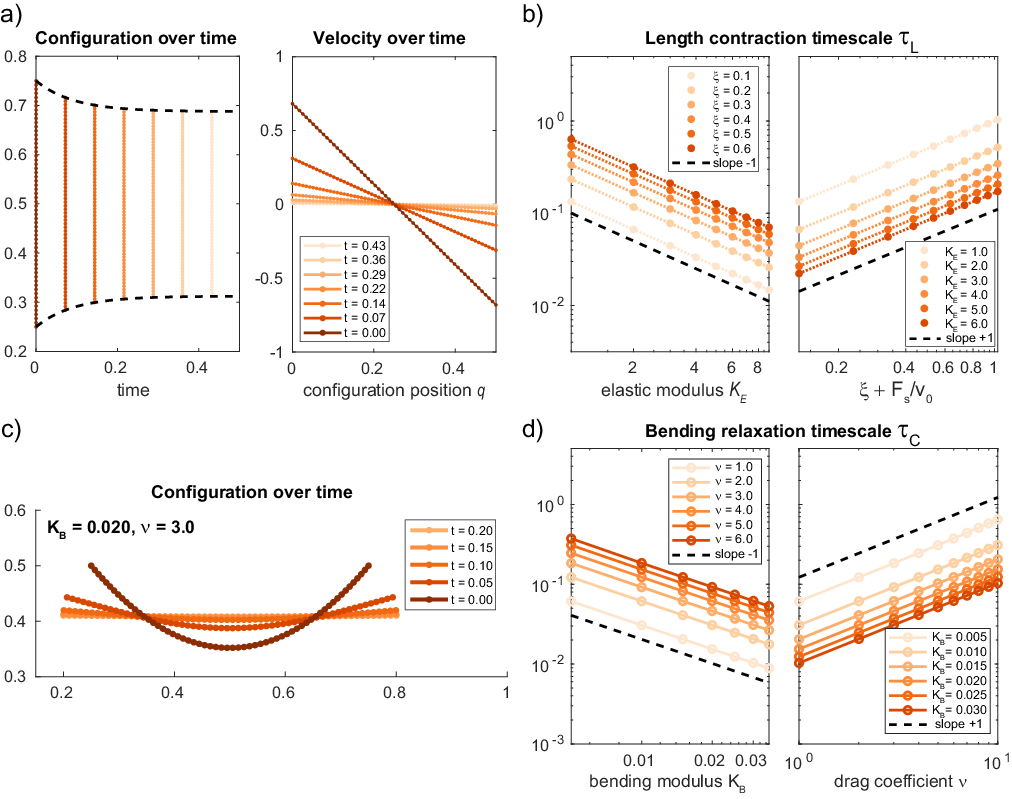}
\caption{\textbf{(a)} The configuration (left) and velocity (right) of a contracting fiber over time, for the case of $K_E=4.0$, $\xi=0.3$, $F_s=1.0$, and $v_0=15$, where the predicted timescale of length contraction is $\tau_L\approx 0.092$. The numerical solution was computed with a $\Delta t = 0.0001$ timestep and $N_b=60$ Lagrangian points. \textbf{(b)} The numerically calculated timescales (solid dots) of length contraction $\tau_L$ for a contracting viscoelastic fiber initialized at rest length $0.5$ for variable elastic modulus $K_E$ (left) and variable viscous resistance $\xi$ (right), with fixed contraction parameters $F_s=0.5$ and $v_0=15$. Dotted lines correspond to the theoretical prediction \eqref{eq:tauL}. The timescales were calculated by numerically solving for the fiber configurations over a $T=1.0$ time with a $\Delta t=0.0001$ timestep and $N_b=60$ Lagrangian points. The length over time was then fit to an exponential using {\tt MATLAB}'s {\tt fit} function. \textbf{(c)}
The configuration of an initially bent, roughly inextensible fiber of nondimensional length $0.6$ over time, for the case of $K_E=100.0$, $K_B=0.02$, and $\gamma=3.0$. The configuration was solved numerically with a $\Delta t = 0.000001$ timestep and $N_b=60$ Lagrangian points. \textbf{(d)} The numerically calculated timescales of curvature relaxation $\tau_C$ for a nearly inextensible fiber ($K_E=100$) under constant linear drag initialized bent at rest length $0.6$ for variable bending modulus $K_B$ (left) and variable drag coefficient $\nu$ (right). The timescales were calculated by numerically solving for the fiber configurations over a $T=0.8$ time with a $\Delta t=0.000001$ timestep and $N_b=60$ Lagrangian points. The maximum curvature over time was then fit to an exponential using {\tt MATLAB}'s {\tt fit} function.}
\label{fig:DragModel_ContractileOrBendingFiber}
\end{center}
\end{figure}

We verify and visualize the timescale's dependence on parameters by looking at the dynamics of a SF which is initialized at rest length ($L_0=0.5$) and allowed to contract until the elastic forces are in balance with contractile forces ($L_\infty=F_sL_0/K_E$). This is effectively the dynamics of the length of a SF after laser ablation, when the fiber has no focal adhesions. Fig.~\ref{fig:DragModel_ContractileOrBendingFiber}(a) shows the length and velocity of a contracting fiber over time where the predicted timescale of length contraction is $\tau_L\approx 0.092$. The length contraction of the fiber matches the timescale as expected, and we see a telescopic velocity profile where the velocity along the length of the fiber increases linearly with distance from the midpoint, as expected for an isolated viscoelastic contractile rod. For fixed $\xi$, $F_s$, and $v_0$, the left plot of Fig.~\ref{fig:DragModel_ContractileOrBendingFiber}(b) confirms the proportionality of the timescale $\tau_L\propto 1/K_E$, as we see that as the elastic modulus $K_E$ increases, the timescale decreases with slope $-1$ on a log-log scale. Similarly, the right plot of Fig.~\ref{fig:DragModel_ContractileOrBendingFiber}(b) confirms how $\tau_L\propto \xi+F_s/v_0$, for as $\xi$ is increased, we see the timescale increase linearly with $\xi+F_s/v_0$ on a log-log scale. Our computed timescales match the theoretical expectation \eqref{eq:tauL}, as shown by the computed values following the dotted lines in Fig.~\ref{fig:DragModel_ContractileOrBendingFiber}(b) which correspond to the theoretical timescale \eqref{eq:tauL}. It is easy also to compute the retraction along the fiber (not shown here), which is constant along the length as expected for a fiber with no adhesions along its length.\\

Now, suppose instead that there is little change in length, and the resistance to curvature governs the dynamics. If $K_E$ is very large to maintain the fiber at rest length, the fiber behaves approximately as an inextensible rod. To assess the impact of $K_B$, it is necessary to balance bending forces with a drag term, for example a linear drag given by $\nu d\pmb{X}/dt$. Then, the fiber configuration obeys the PDE\footnote{This PDE is with dimensions}
\[
\nu \dfrac{\partial \pmb{X}}{\partial t} = -K_B \dfrac{\partial^4\pmb{X}}{\partial q^4}
\]
Consider for example an infinitely long fiber with initial condition where $\pmb{X}(q) = \br{q,\cos(q)}$. The solution would then be
\beq
\pmb{X}(q,t) = \br{q,\cos(q)\exp\pr{-K_B t/\nu}}
\label{eq:InextensibleFiberEquation}
\eeq
As such, the curvature varies over time like:
\[
\pmb{C}(q,t) = \dfrac{\partial^2\pmb{X}}{\partial q^2} = \br{0,-\cos\pr{q}\exp\pr{-K_B t/\nu}}
\]
In our case, the fiber is of course not infinite, and it has free end boundary conditions. But in general, the equation \eqref{eq:InextensibleFiberEquation} as well as dimensional analysis indicates a timescale of curvature relaxation which should be
\beq
\tau_C \propto \dfrac{\nu L^4}{K_B}
\label{eq:tauC}
\eeq
where $L$ is the fiber length. It is straightforward to demonstrate this is true in general for our case. Consider a fiber which has rest length $L_0=0.6$ and is slightly bend around its center as depicted in the darkest color in Fig.~\ref{fig:DragModel_ContractileOrBendingFiber}(c). Over time, as the color goes from dark to light, Fig.~\ref{fig:DragModel_ContractileOrBendingFiber}(c) shows how the fiber straightens, with the ends of the fiber moving out and down while the center moves up in the domain. If one varies $K_B$, the left plot of Fig.~\ref{fig:DragModel_ContractileOrBendingFiber}(d) shows how increasing the bending modulus decreases the timescale with a slope of approximately $-1$ on a log-log plot, as we would expect for a relationship $\tau_C\propto 1/K_B$. Meanwhile, increasing the drag through $\nu$ increases the timescale linearly as expected, as shown in the right plot of Fig.~\ref{fig:DragModel_ContractileOrBendingFiber}(d). 


\subsection{Stress fibers dynamics in a bulk actin network}
Given our understanding of the dynamics of isolated contractile or bending fibers, we want to consider how immersing these fibers in passive bulk actin networks impacts the dynamics, as well as the induced flows in the bulk network. \\
\begin{figure}
\begin{center}
\includegraphics[width=1.0\linewidth, angle=0]{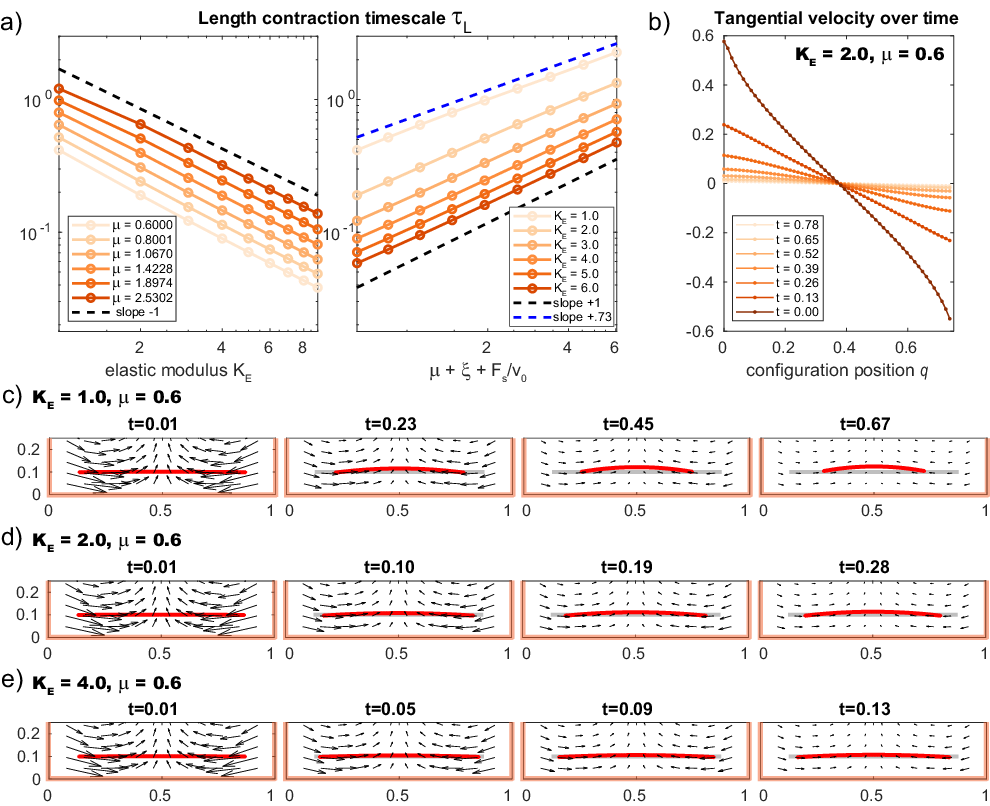}
\caption{\textbf{(a)} The numerically calculated timescales of SF contraction $\tau_L$ for a contracting viscoelastic fiber immersed in a bulk network for variable elastic modulus $K_E$ (left) and variable bulk network viscosity $\mu$, $\lambda=2\mu/3$ (right). The fiber is initialized at rest length $0.75$ at a distance $0.1$ of the edge of the domain from $x=0.125$ to $x=0.875$ with fixed contraction parameters $F_s=0.5$ and $v_0=15$ and internal viscosity coefficient $\xi=0.018$, the largest value for which the fixed point iteration scheme converges for all shown parameter sets. The timescales and subsequent velocity and configuration profiles in \textbf{(b-e)} were calculated by numerically solving for the fiber configurations over a $T=3.0$ time with a $\Delta t=0.001$ timestep, $N_b=60$ Lagrangian points, and on a $100\times100$ actin network Eulerian grid. The length over time was then fit to an exponential using {\tt MATLAB}'s {\tt fit} function. \textbf{(b)} The tangential velocity of a contracting fiber over time for the case of $K_E=2.0$ and a fluid viscosity of $\mu=0.6$. \textbf{(c-e)} The configuration of a contracting fiber over time with elastic moduli $K_E = 1.0,\, 2.0,\,4.0$, immersed in a bulk actin network with viscosity $\mu=0.6$ with the corresponding flow shown in black arrows. The quiver scale is such that a speed of $1$ scales to $0.35$ spatial units. Thick light orange lines mark the boundaries of the domain.}
\label{fig:ActinFluid_ContractileFiber}
\end{center}
\end{figure}

Consider a bulk network which is occupying a $R\times R = 50\mu m\times 50\mu m$ space, and a SF of length $37.5\mu m = 0.75R$ which is also fairly thick, such as $1\mu m$ in diameter\footnote{Since we have observed the hydrodynamic radius of our immersed boundary method to be $O(h)$, for this problem we choose an Eulerian grid which is $h\sim \text{fiber radius}/\text{domain size} = 1/100$. Using $N_b=60$ Lagrangian points, the discretization of the fiber $\Delta q = 0.0125$ is thus appropriately comparable to the fluid with $h=0.01$.}. As SFs are often localized at the cell boundaries \citep{oakes_geometry_2014, ruppel_force_2023}, consider a SF which is a $5\mu m = 0.1R$ distance away from the bulk network boundary. Our aim is to both characterize the dynamics of a contracting SF in this context, as well as determine how the timescale of contraction depends on both the elastic modulus of the fiber $K_E$ as well as the bulk network viscosity $\mu$. As before, we will fix the expansion viscosity to be $\lambda=2\mu/3$, though any expansion viscosity $\lambda = O(\mu)$ is appropriate.\\

Recall that the timescale of length contraction of an isolated fiber \eqref{eq:tauL} predicts an inverse relation with the fiber elastic modulus, $\tau_L\propto 1/K_E$, because as the fiber gets stiffer, it is more resistant to contraction and deforms slower. The left plot of Fig.~\ref{fig:ActinFluid_ContractileFiber}(a) shows that in the case of a fiber contracting in a fluid-like network, this relation continues to approximately hold, as the timescale changes close to a $-1$ slope on a log-log plot as compared to the elastic modulus for various choices of network viscosity $\mu$. For small $K_E$, the timescale can be seen to change a bit slower than with slope $-1$, because when the elastic modulus is small, the network viscous effects start to dominate in their effect on the dynamics and timescale. When it comes to the impact of the bulk network viscosity on the timescale, recall that for an isolated fiber the timescale $\tau_L\propto \xi+F_s/v_0$, where $\xi$ is the internal viscosity of the fiber. As some previous models of SF dynamics under laser ablation consider the cytoplasm as effectively contributing to this internal viscosity \citep{kassianidou_geometry_2017}, we wonder whether the timescale of length contraction varies like $\tau_L\propto \mu + \xi + F_s/v_0$. The right plot of Fig.~\ref{fig:ActinFluid_ContractileFiber}(a) shows that it is not so straightforward, very clearly so for small $K_E$. For larger $K_E=6.0$, the relation between $\tau_L$ and $\mu$ is closer to the prediction, as seen by being close to parallel with a line of slope $+1$ on the log-log plot. Meanwhile for $K_E=1.0$, the relation is approximately $\tau_L\propto \pr{\mu+\xi+F_s/v_0}^{0.73}$.\\

While one might consider this deviation from the expectation to be an effect of proximity of the fiber to the bulk network boundary, these relations between $K_E$ or $\mu$ with $\tau_L$ are conserved if the fiber is positioned differently in the bulk network. Supplemental Fig.~S3(a) shows the same timescale relations for a fiber placed symmetrically in the center of the domain -- the only difference being that the timescales $\tau_L$ have all become slightly faster, as the fiber does not have to compete with stretching flows resulting from contracting near the bulk network boundary. Fig.~\ref{fig:ActinFluid_ContractileFiber}(c-e) shows how while the contractile fiber induces contractile flows along its axis, the placement near the boundary results in bulk network flow perpendicular to the fiber axis, flowing toward the interior of the domain. These flows stretch the fiber, contrary to the contractile action of the myosin along the fiber length, inevitably increasing the contraction timescale due to competing processes. Fibers which have lower elastic moduli $K_E$ become bent as a result, while fibers that are stiffer along their length finish contracting before being hugely impacted, as seen by comparing Fig.~\ref{fig:ActinFluid_ContractileFiber}(c) and Fig.~\ref{fig:ActinFluid_ContractileFiber}(e). Meanwhile, a fiber in the center of the domain contracts along its axis and the flows perpendicular to its axis are outward (see supplemental Fig.~S3), ultimately remaining straight. If the fibers additionally resisted bending, then the length contraction and bending resistance timescales would be in competition, modifying the degree to which a fiber would be bent by contractile flows as shown in Fig.~\ref{fig:ActinFluid_ContractileFiber}(c). \\

Thus, as opposed to the network viscosity simply supplementing the internal viscosity of the fiber, we see that the bulk network is contributing a drag force on the fiber (of the form, for example, $F_{\text{drag}} = \nu d\pmb{X}/dt$). Notice how in an isolated fiber, Fig.~\ref{fig:DragModel_ContractileOrBendingFiber}(a) showed the tangential velocity being perfectly telescopic, but a fiber immersed in a bulk network has a velocity profile, particularly in early time, which increases along the length but not linearly with position, as shown in Fig.~\ref{fig:ActinFluid_ContractileFiber}(b). This is the sort of profile we expect from a drag-like effect of the fluid on the fiber. We expect the bulk network drag forces to be on the scale of \eqref{eq:FluidDragLaw}, but it is evidently not homogeneous throughout the bulk network, as placement of the fiber in different parts of the domain impacts the timescale of length contraction (this is seen by comparing Fig.~\ref{fig:ActinFluid_ContractileFiber}(a) and supplemental Fig. S3(a)). \\
\begin{figure}
\begin{center}
\includegraphics[width=1.0\linewidth, angle=0]{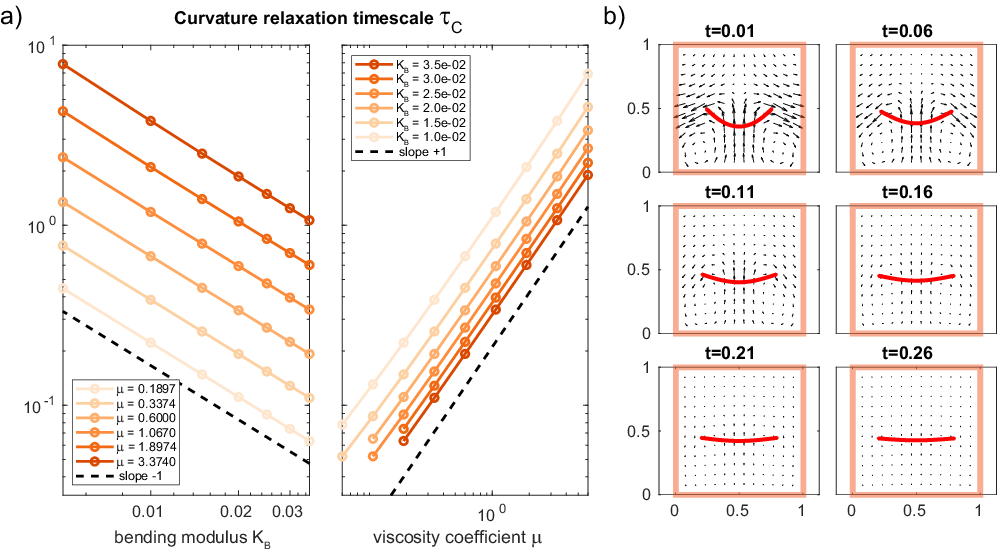}
\caption{\textbf{(a)} The numerically calculated timescales of curvature relaxation $\tau_C$ for a nearly inextensible fiber ($K_E=100$) immersed in a bulk actin network initialized bent at rest length $0.6$ for variable bending modulus $K_B$ (left) and variable bulk network viscosity $\mu$, $\lambda=2\mu/3$ (right). The timescales and subsequent profiles in \textbf{(b)} were calculated by numerically solving for the fiber configurations over a $T=5.0$ time with a $\Delta t=0.00005$ timestep, $N_b=60$ Lagrangian points, and on a $100\times100$ actin network Eulerian grid. Note that omitted combinations of $\mu$ and $K_B$ are those that require a smaller timestep for stability. The maximum curvature of the configuration over time was then fit to an exponential using {\tt MATLAB}'s {\tt fit} function. \textbf{(b)}
The configuration of an initially bent, roughly inextensible fiber over time with elastic and bending moduli $K_E=100.0$ and $K_B=0.02$, immersed in a bulk actin network with viscosity $\mu=0.1897$ with the corresponding flow shown in black arrows. The quiver scale is such that a speed of $1$ scales to $0.2$ spatial units. Thick light orange lines mark the boundaries of the domain}
\label{fig:ActinFluid_BendingFiber}
\end{center}
\end{figure}

Let us now consider the case of a bent, nearly inextensible fiber with the same considered domain size and fiber thickness (recall that the fiber thickness determines the Eulerian grid size and thus fiber discretization). Consider the same initial configuration as in Fig.~\ref{fig:DragModel_ContractileOrBendingFiber}(c), placing the ends of the fiber symmetrically in the center of the domain, as shown in the first time snap of Fig.~\ref{fig:ActinFluid_BendingFiber}(b). For an isolated fiber under constant drag, we saw a timescale which was $\tau_C\propto \nu L^4/K_B$. If we assume that the bulk network imposes a viscous drag on the fiber, as we expected for the case of an immersed contractile fiber, we should see a timescale of curvature relaxation which is $\tau_C\propto \mu L^4/K_B$ (assuming $\lambda=O\pr{\mu}$). Indeed, the left plot of Fig.~\ref{fig:ActinFluid_BendingFiber}(a) shows that the timescale of curvature relaxation decreases approximately like $\tau_C\propto 1/K_B$ for various fluid viscosities, as the timescale plotted against $K_B$ on a log-log plot changes approximately with slope $-1$. Meanwhile the right plot of Fig.~\ref{fig:ActinFluid_BendingFiber}(a) confirms our expectations, where we see that the timescale of curvature relaxation increases approximately proportionally to the bulk network viscosity coefficient $\mu$ for various choices of bending modulus $K_B$. We can also look at the induced flows around the relaxing SF. In general, regardless of viscosity coefficient, Fig.~\ref{fig:ActinFluid_BendingFiber}(b) shows how this initial configuration of the fiber in the bulk network initially induces vortices below it in the domain as the center of the fiber moves upward while the edges of the fiber stretch out. The configuration of the fiber over time is not unlike the isolated case we saw in Fig.~\ref{fig:DragModel_ContractileOrBendingFiber}(c) but the effect on the bulk network is now realized through the use of IBM.\\

The last step of our investigation is to address how myosin activity in bulk networks affects SF dynamics, given that cells' actomyosin networks are usually dynamic \emph{and} contractile.

\subsection{Stress fibers dynamics in contractile actomyosin networks}
Both in micropatterned or scaffolded cells \citep{kassianidou_geometry_2017,brand_tension_2017,vignaud_stress_2021} and free non-motile cells \citep{gavara_relationship_2016, riedel_positioning_2024}, SFs form on the cell periphery, embedded in the actin cortex and lining the cell contour. The cell boundary shape is then effectively a series of concave arcs between major adhesion cites \citep{bischofs_filamentous_2008,schakenraad_mechanical_2020}, resulting from contractility of the cytoskeleton \citep{zand_what_1989,bischofs_filamentous_2008}. There are also SFs in the interior, which can align with the cells major axis \citep{gavara_relationship_2016,schakenraad_mechanical_2020} or run diagonally through to reinforce the cell \citep{riedel_positioning_2024}.\\

\begin{figure}
\begin{center}
\includegraphics[width=1.0\linewidth, angle=0]{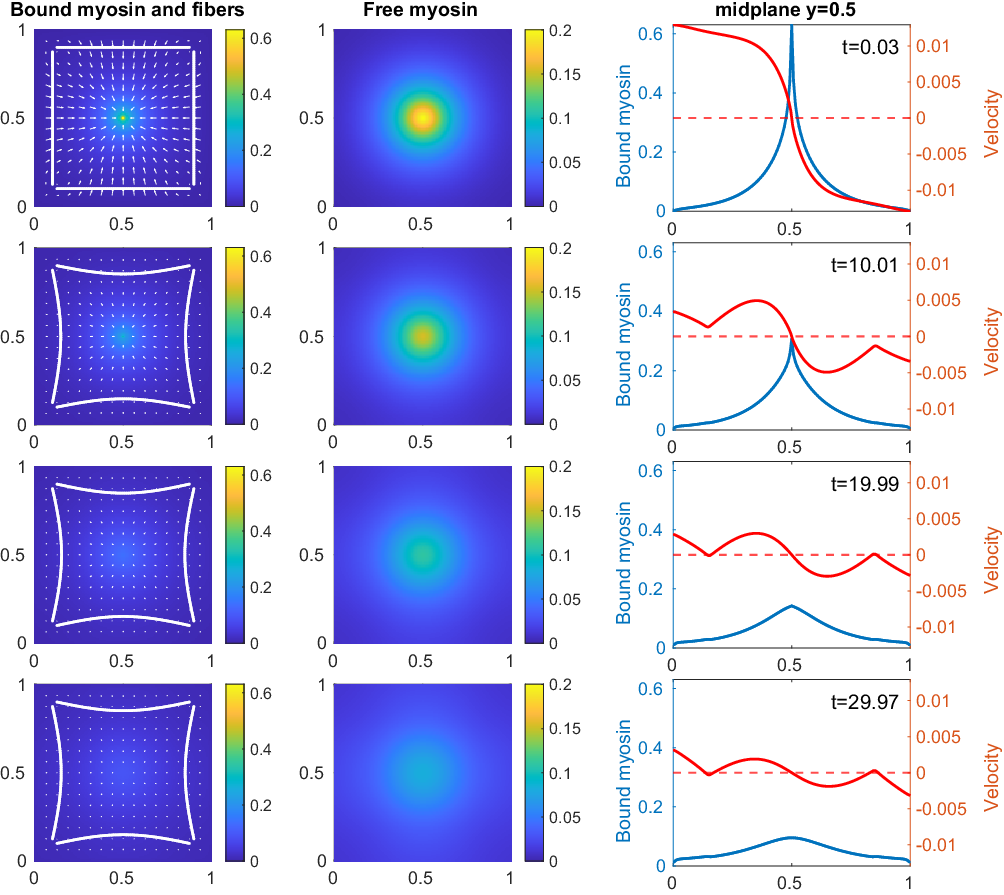}
\caption{The numerical solution over time, before ablation, for the case of four SFs with adhesions at their ends immersed in a bulk actomyosin network with initial bound and free myosin distributions as shown in the top row of Fig.~\ref{fig:MyoFluidParamSweep}(b). The SFs have an elastic modulus $K_E=0.25$, bending modulus $K_B=0.001$, internal viscosity $\xi=0.035$, contractile stall force $F_s=0.05$, and contractile free velocity $v_0=15$. The end adhesions have a spring force of $k_{adh} = 100$. The quiver scale is such that a speed of $1$ scales to $3$ spatial units. The solution is computed over $T=50.0$ time units with a $\Delta t = 0.001$ timestep, $N_b=64$ Lagrangian points, and on a $128\times 128$ actomyosin network Eulerian grid.}
\label{fig:BeforeAblation}
\end{center}
\end{figure}

Let us consider the dynamics of peripheral SFs embedded in a \emph{contractile} actomyosin bulk network with high turnover. Once again, consider a $R\times R = 50\mu m\times 50\mu m$ size bulk network, and suppose we have four SFs of length $37.5\mu m = 0.75R$ which are approximately $0.8\mu m$ thick with focal adhesions at the ends. We initialize these fibers near the edges of the bulk network, specicially $5 \mu m$ away from the boundary, mimicking the effect of having four focal adhesion cites close to the four corners of the domain. For the contractile bulk network, let us consider an initial distribution of free and bound myosin which corresponds to a steady state distribution where the contractile myosin forces are balanced with the network viscosity and exchange with freely diffusing myosin. We choose the case shown in the top row of Fig.~\ref{fig:MyoFluidParamSweep}(b) in this example. However, note that we do not expect this distribution to remain steady in the context of immersed SFs, as they necessarily contribute additional forces and stresses which ultimately will resist flow. The top row of Fig.~\ref{fig:BeforeAblation} shows this initial fiber and network configuration.\\

We expect that the inward flows of the bulk actomyosin network will pull the SFs inward. However, eventually, the adhesions, elastic resistance, contractility, and bending resistance of the fiber should balance the inward flow. Fig.~\ref{fig:BeforeAblation} and Movie 1 show how over time the SFs stretch inward, forming concave arcs, until eventually they become stationary at approximately $T=20$ as can be seen in the rightmost column velocity snapshot. Notably, the right column of Fig.~\ref{fig:BeforeAblation} shows how around the SF, the bulk network still tends to flow inward as a result of the peaked myosin distribution in the center of the domain. As a result of the SFs' resistance to stretch, over time there is a decrease in flow speed of the network, which in turn, due to the exchange of bound myosin to freely diffusing myosin, makes the myosin distribution amplitudes decrease over time. Effectively, the addition of SFs disrupted the balance of contraction and diffusion in the bulk actomyosin network, resulting in a less concentrated myosin distribution over time. In turn, decreased contraction of the bulk actomyosin network allows the SF to relax a little, as seen in the velocity snapshot across $y=0.5$ at time $t=29.97$ in the last row of Fig.~\ref{fig:BeforeAblation} where the velocity switches sign in small regimes around the SFs on the left and right.\\
\begin{figure}
\begin{center}
\includegraphics[width=1.0\linewidth, angle=0]{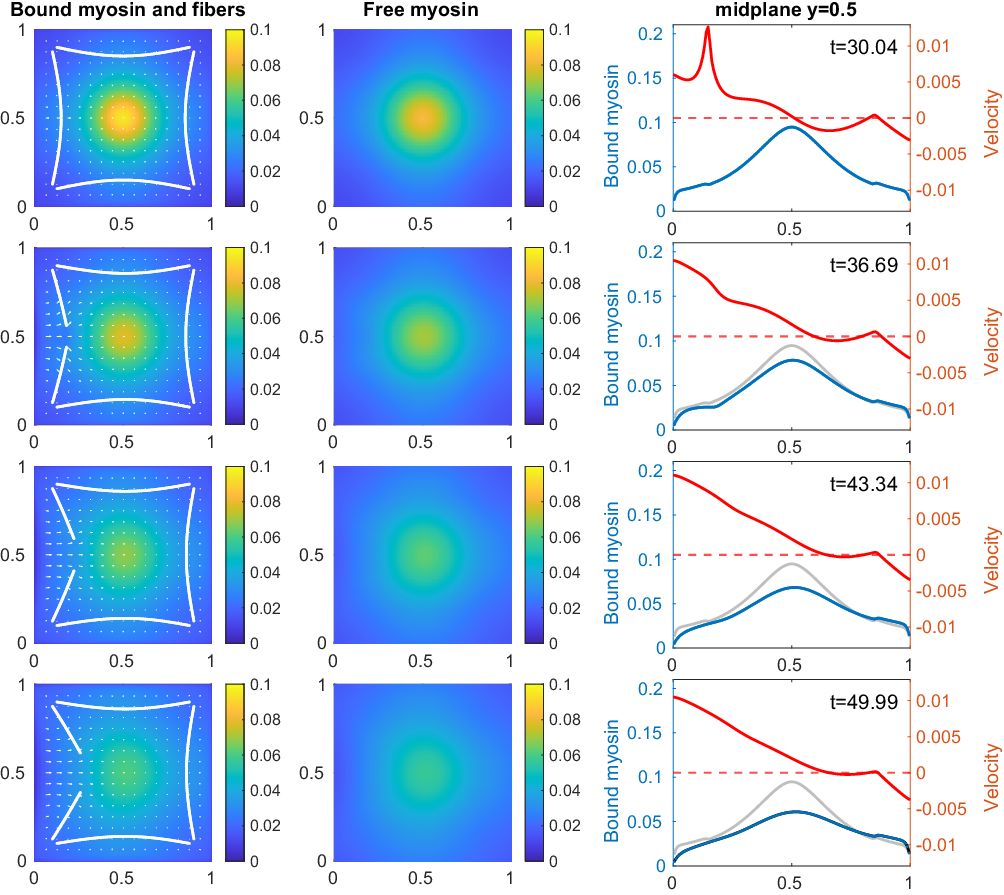}
\caption{Continuing from the final time of Fig.~\ref{fig:BeforeAblation}, the numerical solution over time after ablation of the left SF in the center at time $T=30.0$ until final time $T=50.0$.}
\label{fig:AfterAblation}
\end{center}
\end{figure}

Most studies of the ablation of peripheral SFs utilize mathematical models where the SF is considered in isolation \citep{besser_viscoelastic_2011,chapin_mathematical_2014,kassianidou_geometry_2017,fogelson_actin-myosin_2018}. Only more recently have there been models of embedded SFs \citep{vignaud_stress_2021,riedel_positioning_2024}. However, these models \citep{vignaud_stress_2021,riedel_positioning_2024} consider an \emph{elastic} bulk network, where our model is more applicable in the case of high turnover of the bulk network. In the context of our case study in Fig.~\ref{fig:BeforeAblation}, it is unclear what the dynamics of the SFs would be under ablation.\\

Suppose we ablate the center of the SF located on the left side of the domain. In isolation, an ablated piece of SF should seek to straighten, as well as contract to a final length which is a fraction $\pr{1-F_s/K_E}$ of its initial unstretched length (in this case, a length of approximately $0.75R/2\times(1-F_s/K_E)=0.3R$). However, the fiber still faces contractile flows from the network, and the ablation of the fiber should impact the bulk actomyosin flows as well. Fig.~\ref{fig:AfterAblation} and Movie 2 show the dynamics over time after the left SF in Fig.~\ref{fig:BeforeAblation} was ablated at time $T=30.0$. As a result of the ablation, the resistance to flow of the bulk network between the ablated pieces is gone, allowing for the quick influx of bulk network flow that can be seen in the top-right plot of Fig.~\ref{fig:AfterAblation}. As the cut pieces retract, they simultaneously straighten and are ``pushed aside" by the inward flowing bulk network, resulting in the rotation of the ablated pieces inward into the domain. The flow in the middle of the left boundary of the bulk actomyosin networks quickly returns to a magnitude close to its pre-fiber values (comparing the right column of the third row of Fig.~\ref{fig:AfterAblation} to the right column of the first row of Fig.~\ref{fig:BeforeAblation}). Notably, as a result of the ablation of the SF, left-right symmetry of the actomyosin network is dramatically broken, as the flows are distinctly faster on the left side than the right and the myosin distributions shift toward the right side of the domain.\\

In isolation, the final length of the ablated pieces of SFs is predicted to be $\approx 0.295R$. In the case of fibers in a passive bulk network, they reach this final length with a timescale which is approximately $\tau_L\approx 4.18$ time units (see Fig.~S4 for details). However, as a result of the contractile activity of the bulk actomyosin network, the contraction of the fibers is slowed to a timescale of approximately $\tau_L\approx 3.86$, reaching a larger final length of $\approx 0.307R$, larger than predicted, as shown in Fig.~S4(d). If one is measuring the properties of the SF from such a laser ablation assay, the timescale would inherently suggest the fiber is either weaker, with smaller $K_E$ values, or has more viscous resistance $\xi$ than its true parameters. In addition, while the true ratio of the contractile force and elastic modulus is $F_s/K_E=0.2$, the apparent ratio from the simulated results is instead $F_S/K_E = 0.168$, suggesting the fiber is less contractile than it is in actuality or stiffer. As such, the activity of the bulk actomyosin network changes the apparent quantitative properties of the SFs as measured by a laser ablation experiment. Note, however, that even with these quantiative changes, the SFs exhibit the same qualitative behavior.\\

Fig.~\ref{fig:AfterAblation} also demonstrates how multiple SFs interact hydrodynamically through the bulk actomyosin network. As a result of ablation of the left SF, between $t=30$ and $t=50$ Fig.~\ref{fig:AfterAblation} also shows how \emph{all} the other peripheral SFs, not just the fiber on the other side, relax a little after ablation. In the rightmost plots of Fig.~\ref{fig:AfterAblation}, it is clear that at $t=36.69$ the velocity at the right SF is distinctly positive and nonzero, meaning it is moving outward to the right and relaxing until the forces are again in balance by the final time $t=49.99$. This agrees with experiments which have shown that the laser ablation of SFs can lead to tension release of other parts of the SF network in a cell, or even across cell boundaries \citep{kassianidou_geometry_2017}.

\section{Discussion}\label{Discussion}
In this work, we have utilized a novel approach to the modeling and simulation of SFs embedded in bulk actomyosin networks with high turnover. We consider the case, where bound myosin induces contractile stresses and flows which reinforce myosin compaction, balanced by the network viscous resistance and the unbinding of myosin which then diffuses freely. Through simulations, we characterized the dynamics of such contractile bulk networks in 2D, highlighting the different possible steady states and their dependence on mechanical and kinetic parameters. We then coupled the dynamics of the bulk network with the mechanics of individual SFs through the IBM, allowing to utilize models of SFs which have been extensively studied as isolated viscoelastic structures. Our simulations of individual SFs in passive bulk networks demonstrated the relevance of hydrodynamic effects of both the fiber mechanics on the network as well as the network flows and boundaries on the fiber. Finally, when considering contractile SFs with adhesions embedded in contractile bulk networks, we demonstrated how SFs effectively interact across cell-sized distances as a result of their embedding in the bulk actomyosin network, and how perturbations to SFs can break the symmetry of bulk networks.\\

To our knowledge, this is the first study to consider the simulation of SFs in bulk contractile actomyosin networks with high turnover, particularly through the use of the IBM. The advantage of utilizing the IBM lies in how it couples the SF and active bulk network. We chose here to consider a SF model which corresponds to models of individual SF dynamics under ablation; however, the IBM formulation allows us to easily change the rheology of the SF, and even introduce heterogeneity along the length, as it only requires modification of the force calculations in the Lagrangian fiber framework. We are also able to consider multiple fibers of differing lengths and properties. Unlike previous works, we are thus able through our model to investigate questions as how different contractile higher-order actomyosin structures affect each other.\\

The primary computational constraint on the approach presented here is that the Eulerian grid size of immersed fibers should be on the order of the true radius of the fibers. In 2D, it was sufficient to consider grid sizes $h\sim 1/200 - 1/100$ to simulate fibers of realistic thickness. However, it quickly becomes computationally intractable to consider very thin fibers. Additionally, in the 2D model, our implementation only allows us to consider small values for the internal fiber viscosity when the fiber is embedded in the bulk. However, in 3D we can utilize our modified IBM algorithm \eqref{eq:FiberMotionDiscreteSlip} introduced in full in Appendix~\ref{sec:ModifiedIBM}. Prior work in the Stokes case \citep{maxian_immersed_2020} and our preliminary work for our model indicates that this modified algorithm dramatically reduces the constraints on both Eulerian grid size and internal fiber viscosity. Alternatively, we could implement a semi-implicit method similar to \cite{strychalski_viscoelastic_2012}. The benefit of this computational work in 2D is that we have characterized the effects of the various bulk actomyosin parameters and SF parameters. The 2D case also demonstrates the relevance of bulk flows at large scales, as opposed to small scale flows resolved by fine meshes. So, we are poised to apply our model to investigate the dynamics of SFs in bulk networks in 3D, where we can take advantage of our proposed modified IBM algorithm \eqref{eq:FiberMotionContinuousSlip} and use parameters informed by our work in 2D.\\

In this work, we consider previously published simplified models for SFs, which assume that they are homogeneous along their length and that their properties do not change with time. However, experiments have shown that there is heterogeneity in these properties along their length \citep{costa_buckling_2002,peterson_simultaneous_2004}. Additionally, SFs are dynamic, undergoing reinforcement and repair under tension \citep{burridge_focal_2016}, as well as disassembly or rearrangement on long timescales \citep{vignaud_stress_2021}. In this work, we have also neglected the impact of adhesions along the length of the fibers, but some laser ablation experiments suggest the existence of such adhesions \citep{besser_viscoelastic_2011,kassianidou_geometry_2017}. To include some of these complexities would require further modeling efforts, but in general our formulation allows for the easy modification of the SF rheology so long as one can appropriately discretize the relevant forces on the Lagrangian grid. In future work, in application of this model to study a particular experimental question, we intend to consider different fiber rheologies to determine the appropriate description depending on the problem's timescale and components. In utilizing this model formulation for future work, our characterization of the dynamics and interplay between parameters in lower dimensions will inevitably be informative and instructive in full 3D studies of higher complexity. 

\backmatter

\section*{Statements and Declarations}
\bmhead{Supplementary information}
Please see the supplementary material file for supplementary figures, caption information for Movie 1 and Movie 2, and additional commentary.

\bmhead{Acknowledgements}
We thank O. Maxian, W. Strychalski, and C. Copos for useful suggestions. This work was supported by National Science Foundation grants DMS 1953430 and 2052515.

\bmhead{Competing interests}
The authors declare that they have no competing interests.

\bmhead{Code availability}
The Matlab codes used for solving the 2D partial differential equations for the bulk actomyosin network as well as the implementation of the IBM in 2D for a contractile viscoelastic fiber are available \href{https://github.com/mariyasavinov/ContractileStressFibers.git}{on Github}.

\bmhead{Author contribution}
All authors contributed to the study conception and design as well as methodology. M.S. did the formal analysis and investigation, as well as wrote the original manuscript. C.P. and A.M. edited the manuscript.


\begin{appendices}

\setcounter{figure}{8}
\renewcommand{\thefigure}{\arabic{figure}}

\section{How $K_E$ relates to the Young Modulus $Y$ of the fiber}\label{sec:KEYoungsModulus}
Note that the elastic energy of the fiber \eqref{eq:FiberEnergyElasticContinuous} is in terms of parameter $K_E$ with dimension of force, unlike a traditional spring constant, which has dimension of force over length. To contextualize what $K_E$ represents, suppose that our fiber has some Young's modulus $Y$, which is
\[
Y = \dfrac{\text{stress}}{\text{strain}} = \dfrac{\sigma}{\varepsilon}
\]
The axial stress $\sigma$ is simply the force over the cross-sectional area $A$ of the fiber: $\sigma = F/A$.\\

Suppose then that the fiber is stretched from rest length $L_0$ to new length $L=L_0+\Delta L$, resulting in strain $\varepsilon=\Delta L/L_0$. The magnitude of the force $F$ from this strain is the fiber tension, and our constitutive law \eqref{eq:FiberTensionContinuous} is that tension is a function of the stretch ratio $\nu=\norm{\partial \pmb{X}/\partial q}$:
\[
T(\nu) = K_E\pr{\nu-1}
\]
In the case of a stretched fiber, $\nu=L/L_0=1+\Delta L/L_0$, so the tension, and thus force, is
\[
F = K_E \dfrac{\Delta L}{L}
\]
This yields a direct proportionality between our elastic measure $K_E$ and the fiber Young's modulus:
\[
Y = \dfrac{\sigma}{\varepsilon} = \dfrac{F}{A}\dfrac{1}{\varepsilon} = K_E\dfrac{\Delta L}{L}\dfrac{1}{A}\dfrac{L}{\Delta L} \Longrightarrow K_E = YA
\]
This shows that $K_E$ is a proxy for the Young's modulus; moreover, it demonstrates that $K_E$ is independent of the total length of the fiber $L$. In the context of spatially discretizing the fiber, $K_E$'s independence with respect to length means that one can easily implement a non-uniform discretization along the fiber -- the same $K_E$ ``spring constant" will apply throughout all of the fiber. 
\section{Numerical scheme for bulk actomyosin network model}\label{sec:FluidScheme}
\begin{figure}
\begin{center}
\includegraphics[width=0.45\linewidth, angle=0]{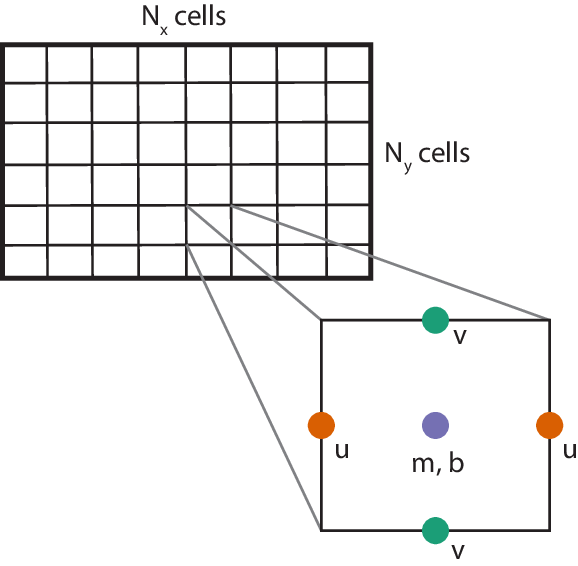}
\caption{Depiction of a standard Arakawa C grid, denoting where velocity $\pmb{u}=<u,v>$ components $u$ (orange) and $v$ (green) are solved, as well as myosin densities $m$ and $b$ (purple).}
\label{fig:ArakawaCgrid}
\end{center}
\end{figure}

We solve the network force balance equation \eqref{eq:FluidForceBalance} and myosin density dynamics equations \eqref{eq:Myosin} with boundary conditions \eqref{eq:FluidBC} and \eqref{eq:MyosinBC} on 2D rectangular domains $\Omega=\br{L_a,L_b}\times\br{L_c,L_d}$ which are of $O(1)$ in size as a result of nondimensionalization by true domain size $R$. We consider an Eulerian grid with $N_x\times N_y$ cells, utilizing a staggered Arakawa C-grid (see Fig.~\ref{fig:ArakawaCgrid}) where the myosin densities $m$ and $b$ are stored at the interior of grids and the fluid velocity components are stored at their corresponding cell edges. \\

We solve the network force balance equation using a second-order finite difference scheme. Recall that we denote the resultant discretized force balance equation as
\[
\pr{\lambda+\mu/3}\pmb{D}\pr{\pmb{D}\cdot\pmb{u}} + L\pmb{u} = -k\pmb{D}m + \zeta\pmb{u} + \pmb{f} \quad\quad\quad \eqref{eq:FluidForceBalanceDiscrete}
\]
where $\pmb{D}$ is the discretized Del operator and $L$ is the discretized Laplacian. The numerical scheme boils down to solving a linear system $Aw=f$, where $A$ is of size
\[
\br{\pr{N_x+2}\pr{N_y+1}+\pr{N_x+1}\pr{N_y+2}}^2\sim 4N_x^2N_y^2
\]
and is banded with upper and lower bandwidth which are approximately
\[2\br{\pr{N_x+2}\pr{N_y+1}+\pr{N_x+1}\pr{N_y+2}}\sim 2N_xN_y\]
In {\tt MATLAB}, we solve this linear system using backslash.\\

The myosin distributions $m$ and $b$ are solved in time at the interior of cells in the Arakawa C-grid with a finite volume approach. The advection-reaction equation in \eqref{eq:Myosin} for bound myosin $m$ is solved using the first-order Corner Transport Upwind (CTU) finite volume scheme, which has stability condition 
\beq
\max\pr{\dfrac{\abs{u}\Delta t}{\Delta x},\dfrac{\abs{v}\Delta t}{\Delta y}} <1
\label{eq:CTUStability}
\eeq
This takes the strictest CFL stability condition between the $x$ and $y$ direction. The free myosin density $b$ obeys a diffusion-reaction equation \eqref{eq:Myosin}, which is solved to second order using the unconditionally stable Crank-Nicolson method.\\

In sum, the bulk network dynamics are solved to first order. We validated our numerical schemes in smooth cases using manufactured solutions for a variety of parameter choices, confirming second or first order convergence of each respective component individually, as well as the first order convergence of the entire scheme. We only see drops in accuracy for the network flow when the nondimensional viscosities $\mu$ and $\lambda$ are very large, reflecting that solutions of \eqref{eq:FluidForceBalance} are only unique \emph{up to a constant} in the absence of myosin forcing or drag terms. In non-smooth cases of myosin distributions, we see a predicted small decrease in convergence rate and evidence of artificial diffusion, which is indeed the primary numerical artifact of the CTU scheme for advection equations. However, additional diffusion is not of particular concern, as in our biological application it effectively contributes to the action of the freely diffusing myosin population. Additionally, achieving several digits of accuracy for small grid sizes is not absolutely necessary, as 1-2 digits of relative accuracy is sufficient in the context of cell biology applications where experimental data is noisy and highly heterogeneous between systems. Overall, we utilize timesteps at or smaller than $10^{-3}$, and thus nearly always operate in the observed asymptotically convergent regimes of our numerical scheme for the bulk actomyosin network.

\section{Spatial discretization of fiber forces}\label{sec:FiberDiscretization}
\subsection{Elastic forces}
We will consider $N_b$ Lagrangian points $\pmb{X}_k$ along the fiber to approximate its configuration $\pmb{X}(q)$. The discretized elastic energy of the fiber \eqref{eq:FiberEnergyElasticContinuous} takes the form
\beq
E_e\br{\pmb{X}_1,...,\pmb{X}_{N_b}} = \dfrac{K_E}{2}\sum_{k=1}^{N_b-1}\pr{\dfrac{\norm{\pmb{X}_{k+1}-\pmb{X}_k}}{\Delta q_{k+1/2}}-1}^2\Delta q_{k+1/2}
\label{eq:FiberEnergyElasticDiscrete}
\eeq
where the energy has become equivalent to the sum of the elastic energies of a series of springs between the Lagrangian points, each with rest length $\Delta q_{k+1/2}$. This has an associated discretized elastic \textit{force} (not force density) at each node $\pmb{X}_k$ given by
\beq
\pmb{F}^e_k= -\dfrac{\partial E_e}{\partial \pmb{X}_k} = T_{k+1/2}\pmb{\tau}_{k+1/2} - T_{k-1/2}\pmb{\tau}_{k-1/2}
\label{eq:FiberForceElasticDiscrete}
\eeq
with tension between nodes $k$ and $k+1$  given by
\beq
T_{k+1/2} = K_E\pr{\dfrac{\norm{\pmb{X}_{k+1}-\pmb{X}_k}}{\Delta q_{k+1/2}}-1}
\label{eq:FiberTensionDiscrete}
\eeq
and unit tangent defined by
\beq
\pmb{\tau}_{k+1/2} = \pr{\pmb{X}_{k+1}-\pmb{X}_k}/\norm{\pmb{X}_{k+1}-\pmb{X}_k}
\label{eq:FiberUnitTangentDiscrete}
\eeq
In the case when the fiber ends are connected, i.e. there is a fiber loop, it is straightforward to adjust the $\pmb{F}_k$ at the ends such as to consider the tension and unit tangent between the first $\pmb{X}_1$ and last point $\pmb{X}_{N_b}$. Otherwise, note that there will be no balance of the tension at the ends.

\subsection{Bending forces}
In the case of a bending fiber, for simplicity we assume that the discretization has all equal reference lengths $\Delta q$. The fiber's bending energy \eqref{eq:FiberEnergyBendingContinuous} can be discretized as
\beq
E_b\br{\pmb{X}_1,...,\pmb{X}_{N_b}} = \dfrac{K_B}{2}\sum_{k=2}^{N_b-1}\norm{\pmb{C}_k}^2\Delta q
\label{eq:FiberEnergyBendingDiscrete}
\eeq
where $\pmb{C}_k$ is the discrete curvature, defined only at interior points:
\beq
\pmb{C}_k = \dfrac{\pmb{X}_{k+1}-2\pmb{X}_k+\pmb{X}_{k-1}}{\pr{\Delta q}^2}\quad\quad k=2,...,N_b-1
\label{eq:FiberCurvatureDiscrete}
\eeq
This energy \eqref{eq:FiberEnergyBendingDiscrete} has an associated discretized elastic force at each node $\pmb{X}_k$ given by
\beq
\pmb{F}^b_k= -\dfrac{\partial E_b}{\partial \pmb{X}_k} = -K_b \dfrac{\pmb{C}_{k-1}-2\pmb{C}_k+\pmb{C}_{k+1}}{\Delta q}
\label{eq:FiberForceBendingDiscrete}
\eeq
At the interior, this is effectively a discretization of the fourth derivative of $\pmb{X}$ with respect to $q$ (except this is \emph{true} force as opposed to force density, so it is missing an additional multiplier $1/\Delta q$). Near the ends of the fiber, the discretization reflects the free-end nature of the fiber. In the case that the fiber ends are connected, i.e. there is a fiber loop, it is straightforward to define curvature at $k=1$ or $k=N_b$ and apply the same discretized force \eqref{eq:FiberForceBendingDiscrete}.

\subsection{Viscous forces}
Recall that in discretizing the elastic fiber, we showed that the total energy is equivalent to the sum of the elastic energy of a series of springs between the Lagrangian points. In discretizing a viscoelastic fiber which behaves like a Kelvin-Voigt solid, we similarly consider instead a series of Kelvin-Voigt elements between the Lagrangian points. Consider the element between the $k$ and $k+1$ Lagrangian points. The viscous force of at $\pmb{X}_k$ due to this element is a discretization of \eqref{eq:FiberForceViscousContinuous} with unit tangent direction $\pmb{\tau}_{k+1/2}$:
\beq
\begin{split}
\pmb{F}^v_k & =\dfrac{\xi}{\Delta q_{k+1/2}} \,\dfrac{\partial}{\partial t}\norm{\pmb{X}_{k+1}-\pmb{X}_k}\,\pmb{\tau}_{k+1/2}\\
& = \dfrac{\xi}{\Delta q_{k+1/2}}\br{\pr{\pmb{U}_{k+1}-\pmb{U}_{k}}\cdot \pmb{\tau}_{k+1/2}}\,\pmb{\tau}_{k+1/2}
\end{split}
\label{eq:FiberForceViscousDiscrete}
\eeq
There is also an analogous force on the Lagrangian point $k$ coming from the element between the $k-1$ and $k$ Lagrangian points.

\subsection{Myosin forces}
We can discretize the myosin force \eqref{eq:FiberForceMyosinContinuous} for each subunit of the fiber, taking the current length as $L\rightarrow \norm{\pmb{X}_{k+1}-\pmb{X}_k}$ and the rest length as $L_0\rightarrow \Delta q$. Consider the element between the $k$ and $k+1$ Lagrangian points. The resultant contractile force due to this element is in the tangent direction $\pmb{\tau}_{k+1/2}$:
\beq
\begin{split}
\pmb{F}_k^m & = F_s\pr{1+\dfrac{1}{v_0 \Delta q_{k+1/2}}\dfrac{d}{dt}\norm{\pmb{X}_{k+1}-\pmb{X}_k}}\pmb{\tau}_{k+1/2}\\
& = F_s\pr{\pmb{\tau}_{k+1/2}-\pmb{\tau}_{k-1/2}}+\dfrac{F_s}{v_0\Delta q_{k+1/2}}\br{\pr{\pmb{U}_{k+1}-\pmb{U}_k}\cdot \pmb{\tau}_{k+1/2}}\pmb{\tau}_{k+1/2}
\label{eq:FiberForceMyosinDiscrete}
\end{split}
\eeq
Note that this is similar to the discretized fiber viscous resistance force \eqref{eq:FiberForceViscousDiscrete}. This is a result of both this myosin force and the viscous resistance depending on the rate of change in the length of a section of the fiber. 

\section{Modified IBM for 3D networks}\label{sec:ModifiedIBM}
Following the approach of \cite{maxian_immersed_2020}, when looking at SFs in 3D bulk networks, we can leverage an analytical result regarding the drag on an sphere immersed in the fluid, instead considering a fiber equation of motion given by
\[
\dfrac{d\pmb{X}}{dt}(q,t) = \pmb{S}^*\pr{\pmb{X}}\pmb{u} + \dfrac{1}{\gamma}\pmb{F}(q)\quad\quad \eqref{eq:FiberMotionContinuousSlip}
\]
One can choose $\gamma$ by solving the equivalent of a Stokes drag in the case of our compressible fluid, meaning calculating the drag force on a sphere of radius $a$ immersed in some background flow $U$. The dimensional actin network force balance equation in this case would be
\[
\pr{\lambda+\mu/3}\nabla\pr{\nabla\cdot\pmb{u}} + \mu\Delta \pmb{u} = \zeta \pmb{u}
\]
Now, non-dimensionalizing this equation by the size of the sphere $a$ and the background flow speed $U$ yields
\[
\dfrac{\lambda+\mu/3}{\zeta a^2}\nabla\pr{\nabla\cdot\pmb{u}} + \dfrac{\mu}{\zeta a^2}\Delta \pmb{u} =  \pmb{u}
\]
As in Sec.~\ref{sec:BulkNetworkNondimensionalization}, we once again identify $\mu/\zeta$ as $\ell_0^2$, the hydrodynamic length squared. Assuming $\lambda$ and $\mu$ are of approximately the same scale, the fluid equation becomes
\[
O(1) \nabla\pr{\nabla\cdot\pmb{u}}+\Delta \pmb{u} = \pr{\dfrac{a}{\ell_0}}^2 \pmb{u}
\]
Note the following: generally, the hydrodynamic length for the cytoplasm is comparable to cell sizes \citep{mayer_anisotropies_2010}. The size of the sphere, determined by radius $a$, in our context would correspond to the thickness of the fiber, which has variable widths often smaller than $0.5\mu m$ \citep{livne_inner_2016} but could potentially be as large as a micron in width \citep{buenaventura_intracellular_2024}. Even in this latter case, $a/\ell_0\sim 10^{-2}$, making the cytosol drag term of the order $O\pr{10^{-4}}$, which is much smaller than the order 1 terms on the left-hand-side. As such, the contribution of the cytosol to the drag on the sphere is approximately negligible, and we can instead solve the problem:
\[
\begin{split}
\pr{\lambda+\mu/3}\nabla\pr{\nabla\cdot\pmb{u}}+\mu\Delta \pmb{u} = 0 \quad& \forall\, r > a\\
\pmb{u}=\pmb{0} \quad & \text{on }r=a\\
\pmb{u}=\pmb{U} \quad & \text{as }r\rightarrow \infty
\end{split}
\]
Due to the inherent symmetry of the problem, let our ansatz for the flow field be that it takes the form
\beq
\pmb{u}\pr{r,\theta} = f(r)\cos\pr{\theta}\pmb{\hat{r}} + g(r)\sin\pr{\theta}\pmb{\hat{\theta}}
\label{eq:DragLawFlowGuess}
\eeq
The specific $f(r)$ and $g(r)$ which satisfy the boundary conditions are
\begin{align*}
    f(r) & = U \dfrac{3\lambda}{\pr{\lambda+10\mu/3}\pr{2\lambda+11\mu/3}} \br{\pr{2\lambda+11\mu/3}-3\pr{\lambda+4\mu/3}\dfrac{a}{r}+\pr{\lambda+\mu/3}\dfrac{a^3}{r^3}}\\
    g(r) & = U \dfrac{3\lambda}{4\pr{\lambda+10\mu/3}\pr{\lambda+11\mu/6}}\br{-2\pr{2\lambda+11\mu/3}+3\pr{\lambda+7\mu/3}\dfrac{a}{r}+\pr{\lambda+\mu/3}\dfrac{a^3}{r^3}}
\end{align*}
Recall that the viscous stress on the fluid is
\[
\pmb{\sigma} = \pr{\lambda-2\mu/3}\pr{\nabla\cdot \pmb{u}}\,\pmb{I} + \mu\pr{\nabla\pmb{u}+\nabla\pmb{u}^T} = \pr{\lambda-2\mu/3}\pr{\nabla\cdot \pmb{u}}\,\pmb{I} + \mu\pmb{\tau}
\]
We integrate the stress over the surface of the sphere to find the resultant drag force. By symmetry, as in the Stokes case, the force should be in the $\pmb{\hat{z}}$ direction:
\begin{align*}
    \pmb{F}\cdot\pmb{\hat{z}} & = \int\int_S \pmb{\hat{r}}\cdot \pmb{\sigma} \cdot\pr{\cos\theta \pmb{\hat{r}}-\sin\theta\pmb{\hat{\theta}}}\,a^2\sin\theta d\theta d\phi\\
    & = 2\pi a^2\int_0^\pi\br{\pr{\mu\tau_{rr}+\pr{\lambda-2\mu/3}\nabla\cdot\pmb{u}}\cos\theta-\pr{\mu\tau_{r\theta}+\pr{\lambda-2\mu/3}\nabla\cdot\pmb{u}}\sin\theta}\sin\theta d\theta\\
    & = \underbrace{2\pi a^2 \mu\int_0^\pi \pr{\tau_{rr}\cos\theta\sin\theta - \tau_{r\theta}\sin^2\theta}d\theta}_{\text{viscous shear part}} + \underbrace{2\pi a^2\pr{\lambda-2\mu/3}\int_0^\pi \nabla\cdot\pmb{u}\pr{\cos\theta-\sin\theta}\sin\theta d\theta}_{\text{viscous expansion part}}
\end{align*}
Notably, we clearly see the contributions to the drag force due to resistance to expansion/compression versus shear. Solving the above integrals, we find the drag force velocity relationship
\beq
F_{\text{drag}} = 108\pi a U \dfrac{\mu\lambda\pr{3\lambda + 4\mu}}{\pr{3\lambda+10\mu}\pr{6\lambda+11\mu}} = 108\pi\eta a U
\label{eq:FluidDragLaw}
\eeq
Notice the similarity with the case of Stokes flow, where $F_{\text{drag}} = 6\pi\mu a U$. Then, in order to choose $\gamma$, suppose we split \eqref{eq:FluidDragLaw} into two parts: an IBM part with the hydrodynamic radius $r_h$ and a Stokes part with a correction radius $r_c$.
\[
\pmb{U} = \dfrac{\pmb{F}}{108\pi\eta r_h} + \dfrac{\pmb{F}}{108\pi \eta r_c}
\]
For equivalency with \eqref{eq:FluidDragLaw}, set $r_c=r_h a/(r_h-a)$. Note that so long as $r_h\geq a$, which is favorable as it means $h$ can be larger than the true radius, then the correction radius is positive as necessary. Identifying the first term on the right-hand-side as the velocity component from the IBM, we recover \eqref{eq:FiberMotionContinuousSlip}:
\[
\pmb{U} = \dfrac{d\pmb{X}}{dt} = \pmb{S}^*\pr{\pmb{X}}\pmb{u} + \dfrac{1}{\gamma}\pmb{F}(q) \quad\quad\text{with}\quad\gamma=108\pi \eta r_c
\]
As expected, the slip coefficient $\gamma$ depends on both viscosity coefficients, though not in a manner which is directly proportional.\\

Recall that $\gamma$ is calculated in the case of an infinite domain -- however, generally we are interested in a finite domain with no-stress boundary conditions on the actin network. In the same manner that we found the environmental drag term is approximately negligible $\zeta \pmb{u}\approx \pmb{0}$ (because the fiber radius is much smaller than the hydrodynamic length), generally the domain size (on the same order as the hydrodynamic length) is much larger than the fiber radius, making contributions from the far field boundary conditions small. As such, so long as the fiber is not too close to the boundary, \eqref{eq:FiberMotionContinuousSlip} approximately holds.

\section{Temporal discretization of immersed boundary numerical scheme}
\subsection{Four-point $\delta_{\Delta x,\Delta y}$ function}\label{sec:4ptDelta}
The four-point $\delta_{\Delta x,\Delta y}$ function is defined as
\beq
\delta_{\Delta x,\Delta y}\pr{\pmb{x}} = \dfrac{1}{\Delta x\Delta y}\phi\pr{\dfrac{x}{\Delta x}}\phi\pr{\dfrac{y}{\Delta y}}
\label{eq:4ptDelta}
\eeq
where $\phi(r)$ is nonzero on the interval $r\in(-2,2)$ and is otherwise zero. It is defined as
\begin{align*}
\phi(r) & = \dfrac{3-2r+\sqrt{1+4r-4r^2}}{8}\quad \text{ on } 0\leq r\leq 1\\
\phi(r-2) & = \dfrac{1}{2}-\phi(r)\\
\phi(r-1) & = \dfrac{1}{2}\pr{r-\dfrac{1}{2}}+\phi(r)\\
\phi(r+1) & = \dfrac{1}{2}\pr{-r+\dfrac{3}{2}}-\phi(r)
\end{align*}
This discretized delta function will effectively ``spread" or ``interpolate" a field (depending on use) to the closest four points in a 1D domain. Note that since we consider a staggered grid, the $x$ component of the elastic force is spread onto the $u$-component grid points and the $y$ component of the elastic force is spread onto the $v$-component grid points, and vice versa for interpolation of the fluid to the fiber.

\subsection{Timestepping in traditional IBM}\label{sec:TraditionalTimestep}
In traditional IBM, immersed structures are typically purely elastic, in which case we can solve for the fiber configuration at each timestep as follows:
\begin{enumerate}
    \item Given a fiber configuration at timestep $n$, denoted as $\pmb{X}^n$, calculate the forces on the fiber:
    \[
    \pmb{F}_k^{n} = \pmb{F}_k^e + \pmb{F}_k^b + \pmb{F}_k^a
    \]
    \item Spread the forces to the fluid:
    \[
    \pmb{f}^n = \pmb{S}\pmb{F}^n
    \]
    \item Solve for the fluid velocity $\pmb{u}^n$ using \eqref{eq:FluidForceBalanceDiscrete}.
    \item Interpolate the fluid velocity to the Lagrangian points:
    \[
    \pmb{U}^n = \pmb{S}^*\pmb{u}^n
    \]
    \item Update the fiber configuration using, for example, a forward Euler timestep:
    \[
    \pmb{X}^{n+1} = \pmb{X}^n + \Delta t \pmb{U}^n
    \]
\end{enumerate}
However, suppose we consider a viscous resistive force $\pmb{F}^v$ or velocity-dependent contractile force $\pmb{F}^m$. Then to calculate the forces in step 1 requires knowing the velocity, which is only solved for in step 4 above. Instead, one can solve for both the configuration $\pmb{X}^n$ \emph{and} the fiber velocity $\pmb{U}^n$ using fixed point iteration. See Appendix~\ref{sec:FPI} for details.

\subsubsection{Timestepping using fixed point iteration}\label{sec:FPI}
We solve the immersed boundary system over time as follows:
\begin{itemize}
    \item \emph{Initialization}:
    At initial time $t_0=0$, we know the fiber configuration, i.e. $\pmb{X}_k^0$ for each Lagrangian point, and the fiber spring rest lengths $\Delta q_{k+1/2}$. Guess that the initial fiber velocity is zero, i.e.
    \[
        \pmb{U}_k^{0,(0)} = 0\quad\quad\forall\,\, k
    \]
    At other timesteps $t_n$, we know the two fiber configurations $\pmb{X}_k^n$ and $\pmb{X}_k^{n-1}$. Guess that the fiber velocity at this timestep is
    \[
        \pmb{U}_k^{n,(0)} = \pmb{U}_k^{n-1} = \dfrac{\pmb{X}^n_k-\pmb{X}^{n-1}_k}{\Delta t}
    \]
    \item \emph{At each FPI step $m=1,2,3,...$}: Given the fiber velocity guess $\pmb{U}_k^{n,(m-1)}$
    \begin{enumerate}
    \item Calculate the forces on the fiber:
    \[
    \begin{split}
        \pmb{F}_k^{n,(m-1)} & = \pmb{F}_k^e +\pmb{F}_k^a + \pmb{F}_k^b +\pmb{F}_k^v\pr{\pmb{U}_{k-1}^{n,(m-1)},\pmb{U}_k^{n,(m-1)},\pmb{U}_{k+1}^{n,(m-1)}} \\ & \quad\quad
        +  \pmb{F}_k^m\pr{\pmb{U}_{k-1}^{n,(m-1)},\pmb{U}_k^{n,(m-1)},\pmb{U}_{k+1}^{n,(m-1)}}
    \end{split}
    \]
    (Note that we omit denoting the dependence of forces on $\pmb{X}_k^n$).
    \item Spread the forces to the fluid
    \[
    \pmb{f}^{n,(m-1)} = \pmb{S}\pmb{F}^{n,(m-1)}
    \]
    \item Solve for the fluid velocity $\pmb{u}^{n,(m-1)}$ using \eqref{eq:FluidForceBalanceDiscrete}.
    \item Take an explicit timestep to find the next approximate fiber configuration and velocity:
    \[
    \dfrac{\pmb{X}^{n+1,(m)}-\pmb{X}^n}{\Delta t} = \pmb{U}^{n,(m)} = \pmb{S}^*\pr{\pmb{X}^n}\pmb{u}^{n,(m-1)} = \pmb{S}^*\pr{\pmb{X}^n}\pmb{u}\pr{\pmb{X}^n,\pmb{U}^{n,(m-1)}}
    \]
    \end{enumerate}
    \item \emph{STOP} iterating when
    \[
    \dfrac{\norm{\pmb{X}_k^{n,(m)}-\pmb{X}_k^{n,(m-1)}}}{\norm{\pmb{X}_k^{n,(m-1)}}} < {\tt tol}
    \]
    for some tolerance {\tt tol}, which we choose to be {\tt tol}$=10^{-5}$. Let this final fiber position be the updated configuration, and set the fiber and fluid velocities at timestep $t_n$ to be
    \[
    \pmb{u}^n = \pmb{u}^{n,(m)}\quad\quad\quad \pmb{U}^n = \pmb{U}^{n,(m)} = \dfrac{\pmb{X}^{n+1,(m)}-\pmb{X}^n}{\Delta t}
    \]
\end{itemize}

\end{appendices}

\bibliography{main_Savinov}

\newpage

\Large
\begin{center} \textbf{Supplement} for \emph{A model for contractile stress fibers embedded in bulk actomyosin networks}
\end{center}
\normalsize

\renewcommand{\thefigure}{S\arabic{figure}}
\renewcommand{\theHfigure}{S\arabic{figure}}
\setcounter{figure}{0}
\setcounter{section}{0}

\section{Bulk actomyosin network dynamics}
\begin{figure}
\begin{center}
\includegraphics[width=1.0\linewidth, angle=0]{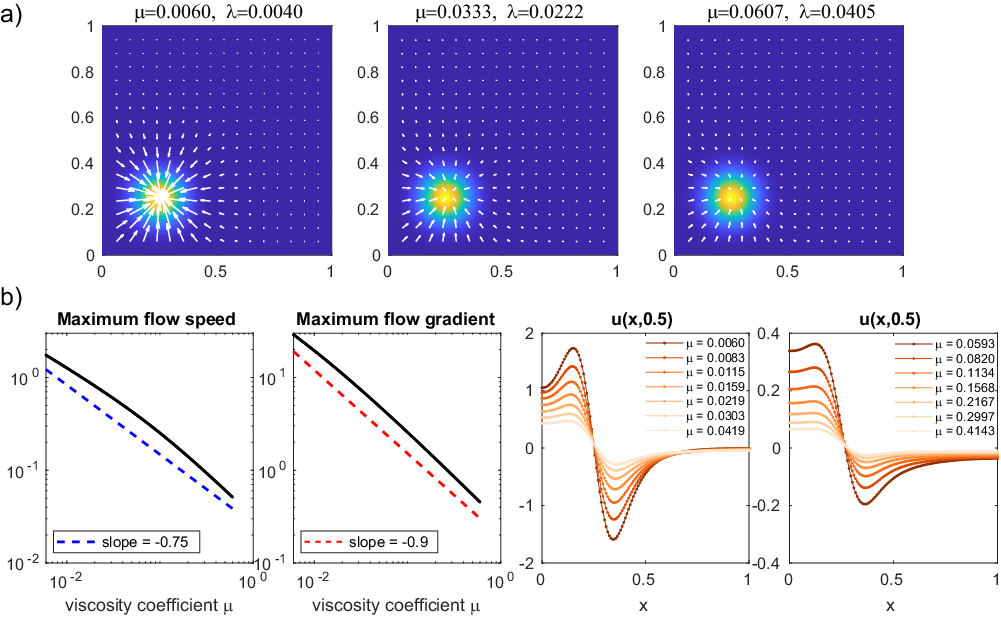}
\caption{\textbf{(a)} Numerical solutions on a $128\times128$ Eulerian grid for the flow of the actomyosin network for different network viscosities in the case of a gaussian-like distribution of bound myosin (31) in the corner $(x_0,y_0)=(0.25,0.25)$ of the domain $\Omega=[0,1]\times[0,1]$. The first viscosity coefficient $\mu$ is varied and the expansion viscosity $\lambda=2\mu/3$ in all cases. The flow quiver arrows are identically scaled between all subplots, such that a flow speed of $1$ appears as an arrow of size $0.07$ in the domain. \textbf{(b)} Quantitative differences in flow properties around the myosin distribution in (a) as a function of changing actin network first viscosity coefficient $\mu$ with expansion viscosity $\lambda=2\mu/3$, as computed on a $128\times128$ Eulerian grid. The left two columns show the maximum flow speeds and gradients, respectively, as a function of increasing viscosity on a log-log plot. The right two columns shows the flow distribution along the horizontal midline of the myosin peak as a function of horizontal position $x$ for a few cases of $\mu$, ranging from $\mu=0.006$ to $\mu=0.4143$, with lighter colors corresponding to larger values of $\mu$.}
\label{fig:FluidParamSweepOffCent}
\end{center}
\end{figure}

As compared to a centered myosin spot as shown in main text Fig 2, if the myosin is instead concentrated near the corner of the domain, Fig.~\ref{fig:FluidParamSweepOffCent} shows that the maximal flow gradient occurring at the center of the spot is mostly unchanged, decreasing like $1/\mu^{0.9}$. However, we see that the maximum flow speed decreases more slowly with $\mu$, more like $1/\mu^{0.75}$. This can be explained by looking at the horizontal flow at $y=0.25$: the no-stress boundary condition results in faster flow of the actin network which is between the spot and the boundary ($0<x<0.25$) than on the other side ($0.25<x<1$). We can explain this physically as follows: there is less actin network for the myosin to pull to the left ($0<x<0.25$) than to the right ($0.25<x<1.0$). The total drag on the left part of the network is smaller than the drag on the right part of the network, so the left part contracts toward the myosin spot faster. Experiments have shown how myosin spots contract corners and edges of actin networks first \citep{colin_friction_2023}, and this agrees with those results.\\

\begin{figure}
\begin{center}
\includegraphics[width=1.0\linewidth, angle=0]{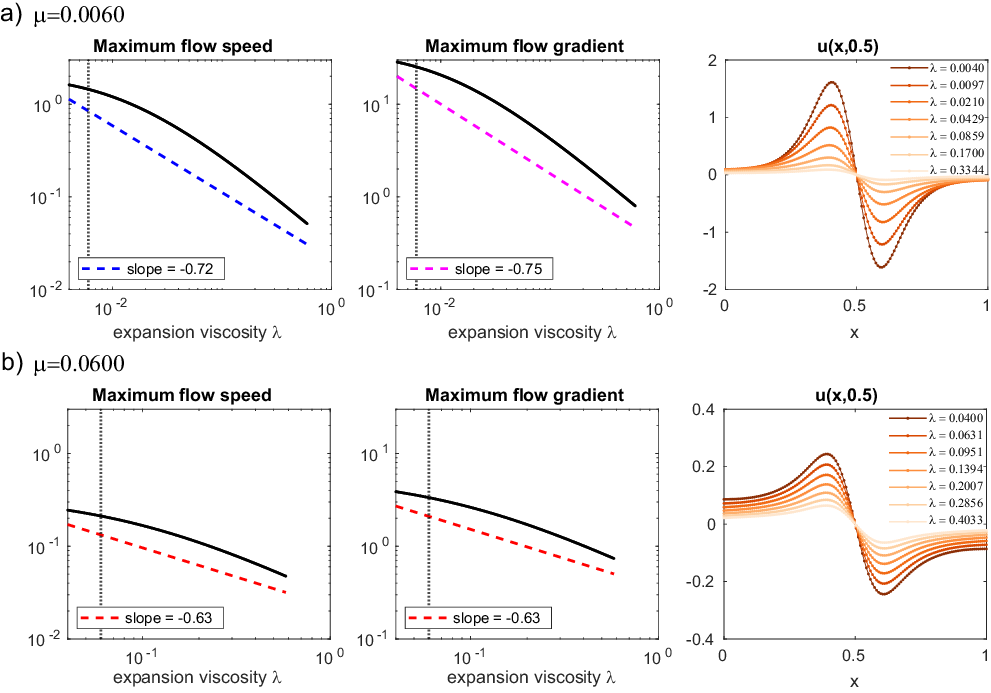}
\caption{Quantitative difference in flow properties around the myosin distribution shown in main text Fig.~2(a) as a function of varying bulk actin network viscosity $\lambda$ for fixed choices of $\mu=0.006$ (top) and $\mu=0.06$ (bottom), as computed on a $128\times128$ Eulerian grid. The left and middle columns show the maximum flow speeds and gradients, respectively, as a function of increasing expansion viscosity on a log-log plot. The right column shows the flow distribution along the horizontal midline of the myosin peak as a function of horizontal position $x$ for a few cases of $\lambda$, starting from $\lambda=2\mu/3$, with lighter colors corresponding to larger values of $\lambda$.}
\label{fig:FluidParamSweepVaryLambda}
\end{center}
\end{figure}

For viscous fluids, the expansion viscosity could in principle be as large as an order or two magnitude larger than the first viscosity coefficient. Meanwhile for water, the first viscosity and expansion viscosity are on the same order of magnitude. Consider fixing $\mu$ and changing $\lambda$: Fig.~\ref{fig:FluidParamSweepVaryLambda} shows the maximum flow speed, maximum flow gradient, and sample flow profiles for the cases of $\mu=0.006$ and $\mu=0.06$. Unlike varying $\mu$ and $\lambda$ together, the change in the flow speed and gradient is farther from a relation like $1/\lambda^p$ for some $p<1$. Similarly to increasing $\mu$, however, increasing the expansion viscosity $\lambda$ results in slower flows, as the actin network is more resistant to the compaction the contractile forces of the myosin spot are driving. Comparing Fig. 2 to plots in Fig.~\ref{fig:FluidParamSweepVaryLambda}, we see that changing $\lambda$ does not result in qualitatively different flow profiles than increasing $\mu$ and $\lambda$ together, aside from exact relation of maximum flow speed to $\lambda$.
\newpage
\newpage
\begin{figure}
\begin{center}
\includegraphics[width=1.0\linewidth, angle=0]{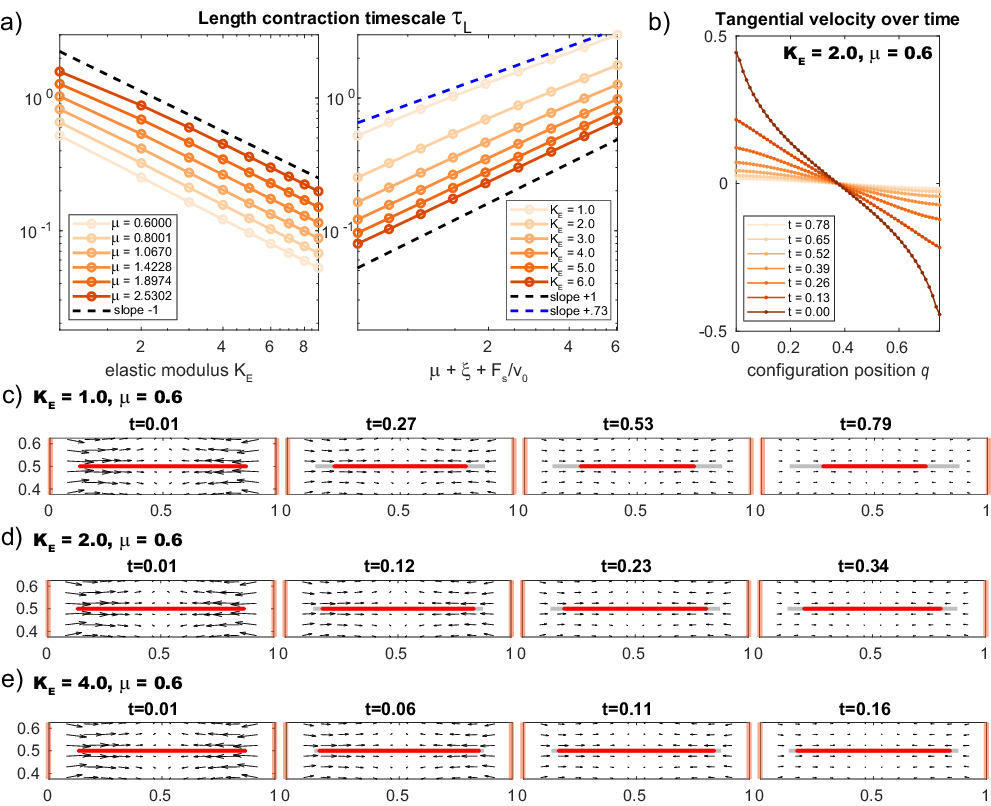}
\caption{\textbf{(a)} The numerically calculated timescales of length contraction $\tau_L$ for a contracting viscoelastic fiber immersed in a bulk actin network for variable elastic modulus $K_E$ (left) and variable bulk network viscosity $\mu$, $\lambda=2\mu/3$ (right). The fiber is initialized at rest length $0.75$ at the center of the domain from $x=0.125$ to $x=0.875$ with fixed contraction parameters $F_s=0.5$ and $v_0=15$ and internal viscosity coefficient $\xi=0.018$, largest value for which FPI converges for all shown parameter sets. The timescales and subsequent velocity and configuration profiles in (b-e) were calculated by numerically solving for the fiber configurations over a $T=3.0$ time with a $\Delta t=0.001$ timestep, $N_b=60$ Lagrangian points, and on a $100\times100$ actin network Eulerian grid. The length over time was then fit to an exponential using {\tt MATLAB}'s {\tt fit} function. \textbf{(b)} The tangential velocity of a contracting fiber over time for the case of $K_E=2.0$ and a fluid viscosity of $\mu=0.6$. \textbf{(c-e)} The configuration of a contracting fiber over time with elastic moduli $K_E = 1.0,\, 2.0,\,4.0$, immersed in a bulk actin network with viscosity $\mu=0.6$ with the corresponding flow shown in black arrows. The quiver scale is such that a speed of $1$ scales to $0.35$ spatial units. Thick light orange lines mark the boundaries of the domain. }
\label{fig:ActinFluid_ContractileFiberCentered}
\end{center}
\end{figure}
\newpage
\begin{figure}
\begin{center}
\includegraphics[width=1.0\linewidth, angle=0]{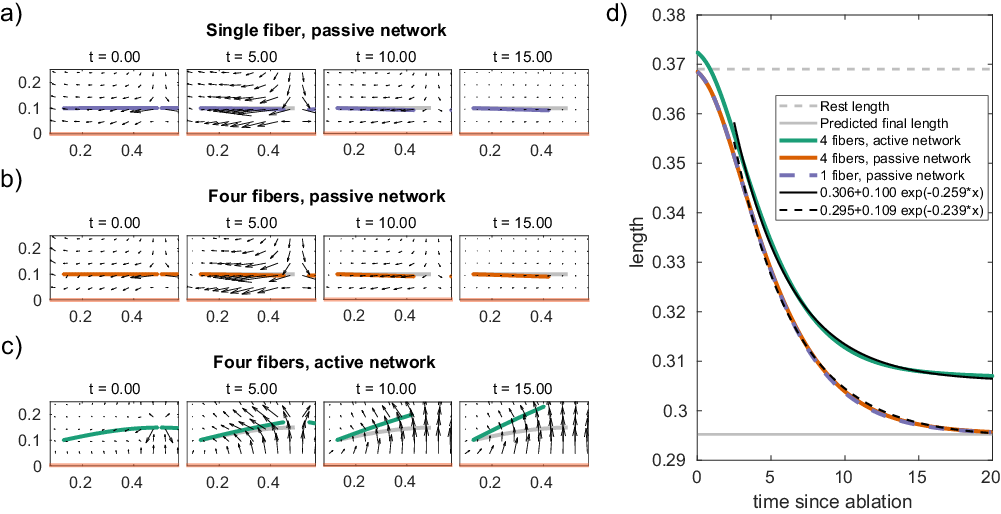}
\caption{The numerical solutions over time for a fiber after laser ablation, in the case of a single fiber in a passive network \textbf{(a)}, four fibers in a passive network \textbf{(b)}, and four fibers in an active network \textbf{(c)}. The length of the piece of fiber after ablation over time in all three cases is shown in \textbf{(d)}, along with exponential fits for the length change after time $T=2.5$. The stress fibers have an elastic modulus $K_E=0.25$, bending modulus $K_B=0.001$, internal viscosity $\xi=0.035$, contractile stall force $F_s=0.05$, and contractile free velocity $v_0=15$. The end adhesions have a spring force of $k_{adh} = 100$. In \textbf{(a)} and \textbf{(b)}, the quiver scale is such that a speed of $1$ scales to $20$ spatial units. In \textbf{(c)}, the quiver scale is such that a speed of $1$ scales to $10$ spatial units. Thick light orange lines mark the boundaries of the domain. The solution is computed over $T=20.0$ time units with a $\Delta t = 0.001$ timestep, $N_b=64$ Lagrangian points, and on a $128\times 128$ actomyosin network Eulerian grid.}
\label{fig:LengthOverTimeAblation}
\end{center}
\end{figure}
\newpage
\section{Captions for Movie 1 and Movie 2}
\textbf{Movie 1}\\
A timelapse of the numerical solution over time, before ablation, for the case of four stress fibers with adhesions at their ends immersed in a bulk actomyosin network with initial bound and free myosin distributions as shown in the top row of main text Fig. 3(b). The stress fibers have an elastic modulus $K_E=0.25$, bending modulus $K_B=0.001$, internal viscosity $\xi=0.035$, contractile stall force $F_s=0.05$, and contractile free velocity $v_0=15$. The end adhesions have a spring force of $k_{adh} = 100$. The quiver scale is such that a speed of $1$ scales to $3$ spatial units. The solution is computed over $T=50.0$ time with a $\Delta t = 0.001$ timestep, $N_b=64$ Lagrangian points, and on a $128\times 128$ actomyosin network Eulerian grid.\\
\textbf{Movie 2}\\
A timelapse of the numerical solution over time starting at time $t=25$, shortly before ablation at time $t=30$, ending at time $t=50.0$ for the same case as Movie 1. The quiver scale is such that a speed of $1$ scales to $3$ spatial units.

\end{document}